\documentclass[
  aip,
  amsmath,amssymb,
  reprint,
]{revtex4-1}

\usepackage{acronym}
\usepackage{graphicx}
\usepackage{dcolumn}
\usepackage{bm}
\usepackage[utf8]{inputenc}
\usepackage[T1]{fontenc}
\usepackage{mathptmx}
\usepackage{etoolbox}
\usepackage[utf8]{inputenc}
\usepackage[T1]{fontenc}
\usepackage{comment}
\usepackage{amsthm}
\usepackage{svg}
\usepackage{enumitem}
\usepackage{miller}
\usepackage{hyperref}
\theoremstyle{plain}

\theoremstyle{definition}

\theoremstyle{remark}
\newtheorem{remark}{Remark}
\usepackage[justification=justified,format=plain]{caption}
\usepackage{amssymb}
\usepackage{amsfonts} 
\usepackage{amsmath} 
\usepackage{url}
\usepackage{booktabs} 
\newcommand{\bs}{\boldsymbol}
\newcommand{\mbf}{\mathbf}
\newcommand{\mbb}{\mathbb}
\newcommand{\mcl}{\mathcal}
\newcommand{\mrm}{\mathrm}

\usepackage{xstring}

\def\R{\mathbb R}
\def\C{\mathbb C}

\def\defeq{:=}

\newcommand*{\norm}[1]{\left\|#1\right\|}

\newcommand*{\card}[1]{\left|#1\right|}
\newcommand*{\diag}[1]{\text{diag}\left(#1\right)}
\newcommand*{\vct}[1]{\text{vec}\left(#1\right)}

\newcommand{\OM}[1]{\textcolor{black}{#1}}%blue

\newcommand{\ABnew}[1]{\textcolor{black}{#1}}%brown
\newcommand{\BM}[1]{\textcolor{black}{#1}}% 
\newcommand{\BMnew}[1]{\textcolor{black}{#1}}% 

\newcommand{\trace}{\operatorname{tr}}

\draft % marks overfull lines with a black rule on the right

\begin{document}
%\preprint{}
% Oleh: Remove Matrix in the title
\title{Eigenstructure Analysis of Bloch Wave and Multislice Formulations for Dynamical Scattering in Transmission Electron Microscopy} 

\author{Arya Bangun}
\email{a.bangun@fz-juelich.de}  % optional, for corresponding author
\affiliation{Forschungszentrum Jülich, IAS 8, Jülich, 52048, Germany}

\author{Oleh Melnyk}
\affiliation{Technical University Berlin, Department of Mathematics, Berlin, 10623, Germany}

\author{Benjamin März}
\affiliation{Louisiana State University, Shared Instrumentation Facility,
Baton Rouge, LA, 70803, United States}

%\affiliation[*]{a.bangun@fz-juelich.de}
 
%\keywords{Keyword1, Keyword2, Keyword3}

\begin{abstract}
We investigate the eigenstructure of matrix formulations used for modeling scattering processes within materials in transmission electron microscopy. 
Dynamical scattering is crucial for describing the interaction between an electron wave and the material under investigation. Unlike the Bloch wave formulation, which defines the transmission function via the scattering matrix, the traditional multislice method is lacking a pure transmission function due to the entanglement of electron waves with the propagation function. To address this, we reformulate the multislice method into a matrix framework, which we refer to as the transmission matrix. This allows a direct comparison to the scattering matrix derived from Bloch waves in terms of their eigenstructures. %We theoretically show their equivalence, given that eigenvalue angles differ by modulo $2\pi n$ (integer $n$), with eigenvectors related by a two-dimensional Fourier matrix. 
Through theory, we demonstrate their equivalence with eigenvectors related by a two-dimensional Fourier matrix, given that the eigenvalue angles differ by modulo $2\pi n$ (integer $n$). We numerically verify our findings as well as demonstrate the application of the eigenstructure for the estimation of the mean inner potential.
\end{abstract} 

\flushbottom
\maketitle
\acrodef{PIE}{Ptychographic Iterative Engine}
\acrodef{ADMM}{Alternating Direction Method of Multipliers}
\acrodef{GS}{Gerchberg Saxton}
\acrodef{HIO}{Hybrid Input Output}
\acrodef{TEM}{transmission electron microscopy}
\acrodef{SM}{Scattering Matrix}
\acrodef{STEM}{scanning transmission electron microscopy}
\acrodef{4D-STEM}{four-dimensional scanning transmission electron microscopy}
\acrodef{MoS$_2$}{Molybdenum disulfide}
\acrodef{GaAs}{Gallium arsenide}
\acrodef{SrTiO$_3$}{Strontium titanate}
\acrodef{MIP}{mean inner potential}
\acrodef{MSE}{mean squared error}
\section{Introduction}\label{sec:introduction}
Modeling the three-dimensional atomic structure of a material is essential for studying its physical characteristics, like atomic electrostatic potential \cite{mendis2025physical, doberstein2025lattice, mendis2024modelling}. Information about the structure can significantly aid in the process of structure retrieval from experimental data such as diffraction patterns, when analyzing the material in transmission electron microscopes \cite{Diederichs2024}. Determining the crystalline structure of materials from diffraction experiments is inseparable from the theory of multiple scattering of electrons, developed by Bethe \cite{bethe1928theorie}. Since then, numerous formulations have been developed to model multiple scattering phenomena in crystals, including the multislice method \cite{cowley1957scattering,ishizuka1977new, dyck1980fast, goodman1974numerical} and the scattering matrix obtained from Bloch waves \cite{sturkey1962calculation, fujimoto1959dynamical}. Both models are frequently used for model-based reconstruction in \acf{TEM} \cite{allen1999retrieval,spence1999dynamic,maiden2012ptychographic,brown2018structure,donatelli2020inversion,findlay2021scattering,chen2021electron,bangun2022inverse,sadri2023determining}.

\begin{figure}[htb!]
   \centering
    \includegraphics[width=\columnwidth]{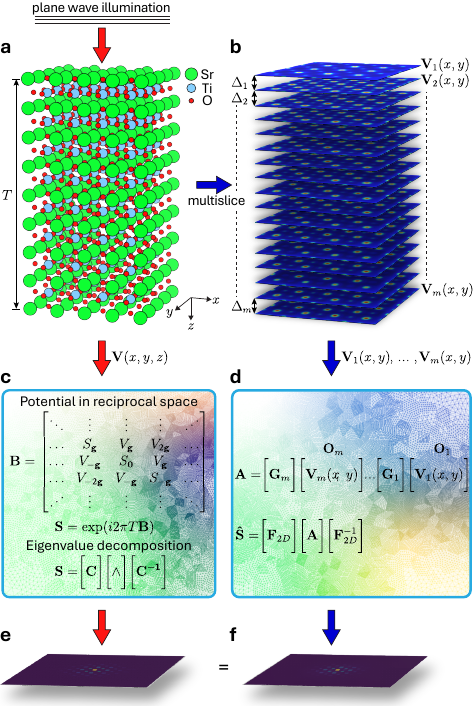}
    \caption{Comparison of the Bloch wave and the multislice approach. \textbf{a}, visualization of a SrTiO$_3$ crystal. \textbf{b}, projected potential representation of the SrTiO$_3$ crystal in \textbf{a} used in the multislice method. \textbf{c}, the construction of the respective scattering matrix $\mbf{S}$ and, \textbf{d}, transmission matrix $\mbf{\hat S}$. 
    % The Bloch wave {method} incorporates all lattice potentials {$V(x,y,z)$ within the specimen of thickness $T$} in a single matrix entity, while {with the} multislice {method} the calculation {is performed} on projected lattice potentials {$V_m(x,y)$} of many thin slices {with distance $\Delta_m$} in several matrices {$\mbf{O}_m$}. %The conventional methods 
    \textbf{e}, \textbf{f}, diffraction patterns generated from matrices $\mbf{S}$ and $\mbf{\hat S}$ using plane wave illumination are similar, yet do not explicitly characterize the differences in eigenstructure of matrices in \textbf{c} and \textbf{d}.} %Comparison of 3D modeling using (a) multislice and (b) Bloch wave formulation as well as the matrix construction. The Bloch wave model incorporates all lattice potentials in a single matrix entity, while multislice performs the calculation on projected lattice potentials of many thin slices in several matrices.
\label{fig:blochwave_vs_multislice}
\end{figure}

Formulating the scattering matrix from the Bloch wave method incorporates the superposition of many periodic waves within a crystal to solve the Schrödinger equation. Therefore, the relation between electrostatic potential and Bloch waves within the crystal can be represented by an eigenvalue problem. Consequently, the scattering matrix is described by its eigenvalue decomposition, with the eigenvalues encoding parameters such as the specimen’s thickness, and the eigenvectors providing the Fourier coefficients of Bloch waves. %Consequently, the formulation of the scattering matrix can be described as an eigenvalue decomposition, with the eigenvalues forming a diagonal matrix containing parameters such as the specimen’s thickness and the eigenvectors as the Fourier coefficients of Bloch waves. 
On the other hand, the multislice method approximates the total electrostatic potential of the crystal by subdividing it into many thin slices of projected potentials. Then, the incident electron wave interacts with all slices consecutively, with the exit wave of one slice being used as the input wave for the following. A visualization of the Bloch wave and multislice approaches is given in Fig.~\ref{fig:blochwave_vs_multislice}.
An important benefit of the multislice method is that the eigenvalue decomposition can be avoided, facilitating a more efficient computation of the propagation through the crystal.

The relationship between Bloch waves and the multislice method is discussed, for instance, in \cite{de2003introduction}. %The idea stems from the fact that the scattering matrix is expressed in terms of matrix exponential of the crystal structure matrix.
In several publications \cite{yang2017quantitative, self1983practical, koch2000comparison, durham2022accurate}, both models are numerically compared. Apart from the computation time, these analyses are mainly focused on the error in the intensity of the specimen exit wave functions for both methods, as this is the quantity that is finally obtained on a \ac{TEM} detector.

While such comparisons are valid to capture the performance, accuracy and consistency of outputs produced by both approaches, the underlying model is treated rather as a black box system and similarity is measured only implicitly.
%\OM{%SUGGESTION: 
%More importantly, measuring similarity in intensity does not provide information about the structure of atom potentials in the Bloch wave or multislice approach.} 
% since the phase information is lost
More importantly, measuring similarity solely based on intensity captures the projection information, which does not provide insight into the %3D
spatial structure of the electrostatic potentials, neither by the Bloch wave nor the multislice approach. %To avoid this, model comparisons should directly evaluate the 3D atomic model rather than relying on intensity alone.
In our study, we therefore go a step further and, rather than relying on the projected intensity alone, directly evaluate the 3D electrostatic potential.
%Additionally, it also \AB{results in} losing the information and structure of atom potentials in the Bloch wave or multislice approach since the phase information is lost. 
%Evaluating only the amplitude and phase of the exit wave from both Bloch wave and multislice methods has limitations, as the atomic potential information is entangled with the input wave that illuminates the material. 

To compare the models, we study how both methods transform the incidence wave. Using Bloch waves, we have a direct expression in terms of the Fourier coefficients; such an expression can not be directly achieved for the multislice method.
%In the Bloch wave, the scattering matrix is expressed in terms of the exponential of the crystal structure matrix.  
% The latter is decomposed into a product of phase grating terms and Fresnel propagation operators. However, the challenge in separating them lies in the non-commutative nature of the matrix product, making a simple decomposition valid only by applying the first-order Taylor approximation. This is the core idea of the multislice method. 
% Although the derivation of the multislice algorithm is straightforward, it is difficult to quantify any differences to the Bloch wave method when modeling scattering processes in crystalline materials.}
In the original formulation of the multislice method, the possibility of defining a pure transmission function, akin to the scattering matrix in the Bloch wave approach, is precluded due to the inherent entanglement between the exit wave across slices and Fresnel propagation. As a consequence, a direct comparison of the eigenstructure of the scattering matrix with the multislice method is made impossible. To circumvent this limitation, we reformulate the multislice model using matrix products and achieve a crucial separation between the input wave and the matrix that represents the transmission function of the crystal structure, which we call \textit{transmission matrix}. This reformulation enables us to model the transmission function in a manner directly comparable to the scattering matrix by using eigenstructure analysis. Rather than the conventional approach of comparing resulting intensities, we present a direct mathematical comparison of the transmission matrix with the scattering matrix.
Furthermore, we investigate whether the similarity between both matrices implies that also the underlying atomic structure is the same by evaluating diffraction patterns simulated by both methods. In addition to that, we leverage the determinant (the product of eigenvalues) of the transmission matrix to estimate the \ac{MIP}, an important physical property of the inner structure of a crystal often analyzed in \ac{TEM} and particularly electron holography experiments \cite{Winkler2016,Lorenzen2024}.

At first, we briefly introduce the scattering and transmission matrices:

\subsection{Transmission Matrix.}
Our formulation of the transmission matrix is derived from the multislice method. Instead of calculating an analytical solution, such as in the Bloch wave approach, the transmission function is built step-by-step using infinitely thin `sub-specimens' (slices) represented by $N \times N$ matrices. The (real-space) transmission matrix of a crystal with $M$ slices is given by
\begin{equation}\label{eq: transmission matrix}
\mathbf{A} = \prod_{m= 1}^{M}\mathbf{G}_{m}\mbf{O}_m,     
\end{equation}
where $\mbf{O}_m \in \C^{N^2 \times N^2}$ is the diagonal matrix representing the potential of each slice, and $\mbf G_m$ is the Fresnel propagator in free space. The elements of both potential and Fresnel propagator matrices depend on interaction constants, spatial frequencies, and tilt angles of the incoming wave. As scattering is typically considered in terms of the Fourier coefficients, so-called reciprocal space, we transform equation \eqref{eq: transmission matrix} as 
% describes the scattering in 
% Unlike the scattering matrix, which is defined in reciprocal or Fourier space, the transmission matrix $\mbf{A}$ operates in real or spatial coordinate space. \BMnew{For this reason, we instead consider}%Instead, we consider:
\begin{equation}\label{eq: multislice scattering reciprocal1}
\mbf{\hat S} = \mbf{F}_{2D} \mbf{A} \mbf{F}_{2D}^{-1},    
\end{equation}
where $\mbf{F}_{2D}$ and $\mbf{F}_{2D}^{-1}$ denote the two-dimensional Fourier and inverse Fourier transforms, respectively. The outlined transformation models the diffraction pattern in the back focal plane of a TEM.

\subsection{Scattering Matrix.}
The scattering matrix $\mbf{S}$ describes the interaction of an input wave function with a crystal, and it can be used to determine a crystal's transmission properties. The derivation of the scattering matrix in \ac{TEM} is based on the theory of electron scattering in crystals, namely,
%It starts by incorporating the Bloch wave to solve the Schrödinger equation and expands the potential function of crystal with Fourier series, also refer as reciprocal space. 
by incorporating Bloch waves to solve the Schrödinger equation. Expanding Bloch waves into Fourier series 
%The potential function of the crystal, which is also referred to as reciprocal space, is expanded using Fourier series.} 
% Incorporating both the Fourier series and the Bloch wave function to solve the Schrödinger equation 
leads to 
% the formulation of the eigendecomposition of 
the scattering matrix of the form
\begin{equation}
    \label{eq:scatmat_main}
    \mbf{S} := \mbf{C} \bs \Lambda \mbf{C}^{-1} \in \C^{N^2 \times N^2},
\end{equation}
where $\mbf{C}$ is a unitary matrix whose columns are the eigenvectors of $\mbf{S}$, representing the Bloch wave Fourier coefficients, and $\Lambda$ is a diagonal matrix containing the corresponding eigenvalues.
These eigenvalues of modulus one encode phase shifts introduced by scattering in the specimen.
% \OM{Here, the potential is represented by $N \times N \times T$ volume, where the resolution of the grid is $N \times N$ and $T$ is the specimen thickness.}   

% \BMnew{By using the multislice method as alternative to the Bloch wave approach, eigenvalue decomposition can be avoided which facilitates a more efficient computation as Fast Fourier Transformation can be used.} The visualization and construction of scattering and transmission matrices is given in Figure~\ref{fig:blochwave_vs_multislice}.

\section{Results} \label{sec:results}

%In what follows, we discuss the structural similarity between transmission matrix $\mbf{\hat S}$ and the scattering matrix $\mbf{S}$. Namely, the equivalence of both matrices is investigated by leveraging their eigenstructures, i.e., their eigenvalues and eigenvectors.
In this section, we discuss the structural similarity between transmission matrix $\mbf{\hat S}$ and the scattering matrix $\mbf{S}$, which we investigated by leveraging their eigenstructures, i.e., their eigenvalues and eigenvectors. 
% \ABnew{We discuss the structural similarity between a matrix obtained using the Bloch wave formulation in comparison to a matrix realized through the multislice method\BMnew{.} %, which throughout this article we refer to as \textit{scattering matrix} and \textit{transmission matrix}, respectively. %%BENnew: commented out previous sentence, as we mention this in the introduction already.
%
%We follow the construction of {the} transmission matrix, as in \eqref{eq: multislice scattering reciprocal1}%.
% \BMnew{We follow \eqref{eq: multislice scattering reciprocal1} for the construction of {the} transmission matrix $\mbf{\hat S}$},
%In this article, we discuss the structural similarity between scattering matrix from Bloch wave formulation and a matrix constructed by reformulating the multislice method, as discussed in \cite{bangun2022inverse}.
%
%The transmission matrix constructed from the multislice method %removes
% and eliminate the dependency {on,} and the entanglement with, the input wave. %Hence, %the
% \BMnew{This way, }its eigenstructure can be evaluated and compared to \BMnew{that of} the scattering matrix \BMnew{$\mbf{S}$} %from
% {of} the Bloch wave {method}. %In summary, our contribution will be discussed as follows:
%\begin{itemize}
    %\item 
We discuss how this equivalency provides insights into the underlying physical processes of dynamical scattering and the accuracy of the approximations resulting from each method. 
    %{We investigate the conditions on eigenstructure under which scattering matrices derived from Bloch wave and multislice are equivalent. HERE WAS MORE COMMENTED, I AM NOT SURE WHAT YOU WERE TRYING TO SAY}
    % We present a different formulation to investigate the condition to have equivalence between scattering matrix derived from Bloch wave and multislice in terms of their eigenstructure. Additionally we perform evaluation and discuss the condition what should be fulfilled to have the equality, for instance the structure of eigenvalues and eigenvectors of both matrices.
   % \item We analyze the eigenstructure of the {scattering} matrix constructed from the multislice method, and draw the connection with physical parameters such as material thickness and {its} energy potentials. 
    % NOT SURE IF THIS IS RELEVANT HERE
    %This analysis is only possible by reformulating the conventional multislice method into matrix formulation.
    %\item 
    %\item 

We conduct numerical simulations to generate the atomic potentials as well as scattering and transmission matrices for three different ideal crystals (GaAs, SrTiO$_3$, Au), %calculated 
directly from the multislice and Bloch wave models. Both matrices are then compared by analyzing the distribution of the eigenvalues in terms of plane-wave diffraction patterns, which we generate using $\mbf{S}$ and $\mbf{\hat S}$, the estimated projected potentials as well as \ac{MIP}s.

We propose a novel approach for the estimation of the \ac{MIP} facilitating the product of the eigenvalues (i.e. determinant) of the transmission matrix, which we then validate by computing the \ac{MIP} of a MoS$_2$ crystal from an experimental dataset obtained by \ac{4D-STEM}.
\subsection{Characterizing Similarity.}
Comparing the amplitudes of both exit waves after interaction with either the scattering or the transmission matrix is straightforward. However, it does not provide the means to investigate the (projected) potentials of a crystal solely from diffraction patterns. 
%, which are encoded in the phase information. 
We, instead, propose to directly investigate the structural similarity between scattering and transmission matrix in terms of their eigenstructures, as presented in Fig.~\ref{fig:blochwave_vs_multislice}c, d. 

Since the scattering matrix $\mbf{S}$ is directly defined by its eigendecomposition in equation \eqref{eq:scatmat_main}, we perform the eigendecomposition of the transmission matrix $\mbf{\hat S}$, yielding the matrix of eigenvectors $\mbf W$ and the diagonal matrix $\Sigma$ containing eigenvalues. 
%Furthermore,% We can \BMnew{then} derive the structural similarity in terms of Frobenius norm. }
%\OM{OM: What bothers me is why specifically this separation and not just entries are equal? Is there a physical motivation for it?}
%\ABnew{AB: Here I just want to mention that the Eigenvector $\mbf{W}$ of (Real SPace) Transmission matrix $\mbf{A}$ and Eigenvector scattering matrix is related by Fourier transform. $\mbf C = \mbf{F}_{2D} \mbf{W}$,  since $\mbf{C}$ is the Fourier transformed of Blochwave, then $\mbf W$ is Blochwave itself. I think we should highlight this relation and that is the reason i want to separate it. }
The equivalence between the scattering and the transmission matrix holds as long as the following conditions are satisfied:
\begin{itemize}[nosep]
    \item Eigenvectors $\mbf C$ of scattering matrix $\mbf S$ and eigenvectors $\mbf W$ of matrix $\mbf{\hat S}$ are equal to the relation $\mbf W = \mbf{F}_{2D} \mbf{V}$ and  $\mbf{V}$ is eigenvector of real space transmission matrix $\mbf A$; %\ABnew{Not correct. It should be  $\mbf C = \mbf{F}_{2D} \mbf{W}$}
    %related in terms of  a two-dimensional Fourier matrix, namely $\mbf{C} = \mbf{F}_{2D} \mbf{W}$. 
    \item As the transmission matrix is unitary, its eigenvalues also have unit modulus. Thus, the eigenvalues $\Lambda$ and $\Sigma$ coincide if the respective angles differ by up to modulo $2\pi n$ for integers $n$.
    % The angles of {both the eigenvalue matrix} %matrix eigenvalues 
    % $\Lambda${, obtained} from scattering matrix $\mbf S${,} and {the eigenvalue matrix} %matrix eigenvalues 
    % $\mbf{V}${, obtained} from matrix $\mbf A${,} differ by up to modulo $2\pi n$ for integer{s} $n = \gamma_k T - \theta_k$, %\ABnew{where $T$ is the specimen thickness, $\gamma_k$ and $\theta_k$ are the parameter given from both eigenvalues scattering and transmission matrix, respectively.}
    % \BMnew{where $T$ is the specimen thickness, $\gamma_k$ and $\theta_k$ are the parameters of the eigenvalues of scattering and transmission matrix, respectively.}
\end{itemize} 

\begin{figure}[htb!]
    \centering
    \includegraphics[width=\columnwidth]{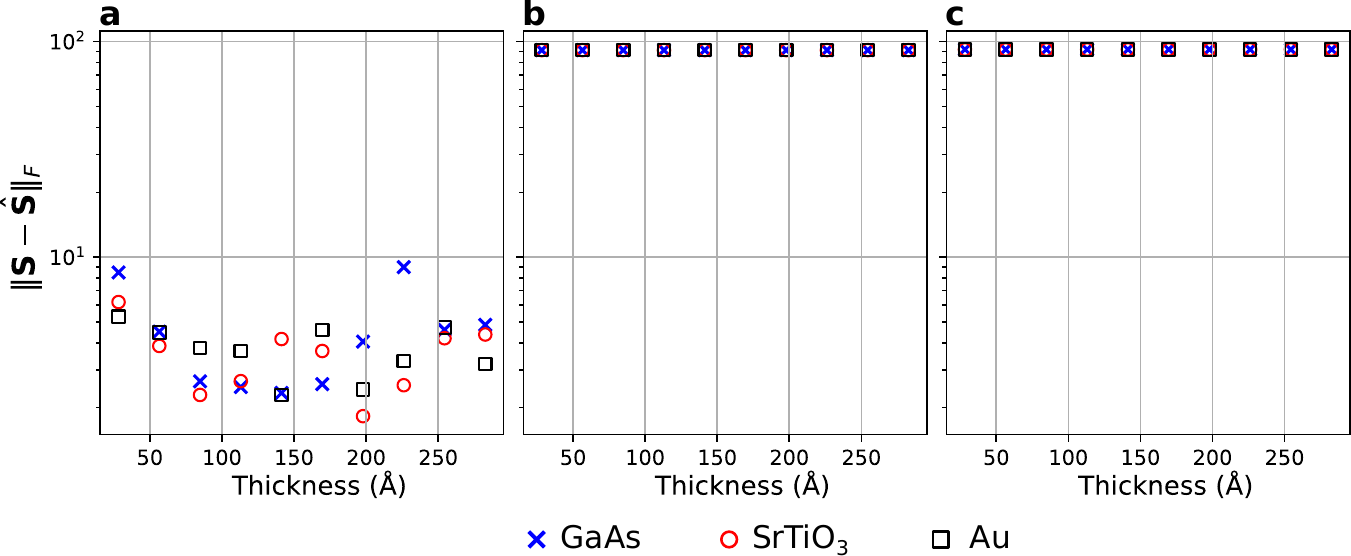}
    \caption{ 
    %Mean squared error between transmission matrix and scattering matrix as a function of crystal thickness of, \textbf{a}, Eigenvalues $\Sigma$ and $\Lambda$ as well as of, \textbf{b}, Eigenvectors $\mbf W$ and $\mbf C$. Results for GaAs, SrTiO$_3$ and Au are shown.
    Frobenius norm as a function of crystal thickness, calculated using the sorted index of eigenvalues and eigenvectors. \textbf{a}, Frobenius norm between eigenvalues $\Sigma$ (transmission matrix) and $\Lambda$ (scattering matrix). \textbf{b}, Frobenius norm between eigenvectors $\mbf W$ (transmission matrix) and $\mbf C$ (scattering matrix). \textbf{c}, Frobenius norm between transmission matrix $\mbf{\hat{S}}$ and scattering matrix $\mbf S$. Results for GaAs, SrTiO$_3$ and Au are plotted.}
    \label{fig:eig_error}
\end{figure}
Given that $\mbf S$ and $\mbf{\hat S}$ are not equal, the difference will affect the eigenvalues and eigenvectors, with the latter being less %stable to errors
robust against errors\cite{kahandavis}.
% \paragraph{Eigenstructure Evaluation}
% A much more precise assessment of the similarity of %both
% transmission and scattering matrix can be achieved by analyzing their eigenstructure, such as eigenvalues and eigenvectors.
% Since both %transmission and scattering matrix 
% are unitary matrices and have eigenvalues with unit modulus, the comparison in terms of amplitude yields no differences. 
Therefore, the first step of our simulation study is to compute the matrices $\mbf S$ and $\mbf{\hat S}$ and their eigen decomposition for GaAs, SrTiO$_3$ and Au for varying specimen thickness.
We depict the Frobenius norm errors between the scattering and the transmission matrix, eigenvalues and eigenvectors in Fig.~\ref{fig:eig_error}. %It shows that $\mbf S$ and $\mbf{\hat S}$ are close to each other with an error of around $10^2$. 
A very similar error of around $10^2$ is shown for $\mbf S$ and $\mbf{\hat S}$. 
Additionally, the error between transmission and scattering matrix is dominated by the error of their eigenvectors, emphasizing the sensitivity of eigenvectors to error. While the corresponding eigenvalues differ with an error of around $0 - 10$, the error between eigenvectors remains constant at $10^2$ within the analyzed thickness range.
\begin{figure}[htb!]
    \centering
\includegraphics[width=\columnwidth]{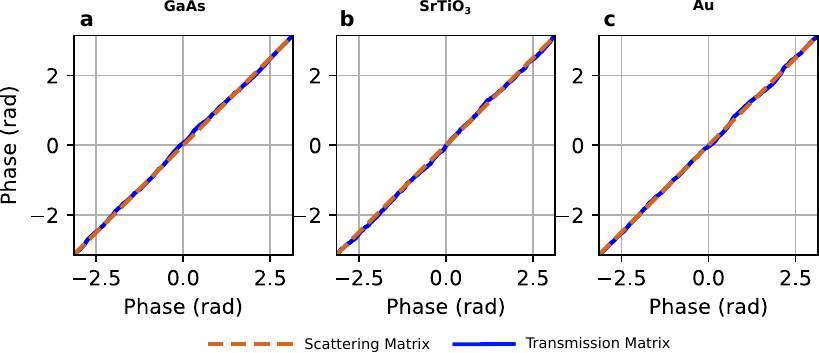}
    \caption{
    The alignment of the eigenvalue angles. \ABnew{Since the total of eigenvalues are the same for both transmission and scattering matrices, both axis represent the value of eigenvalue angles in radian.} The x- and y-coordinates are the eigenvalue angles of the scattering and transmission matrix, respectively. \textbf{a}, GaAs with a thickness of $169.6$\,\AA{} and a standard deviation (SD) of $0.037$\,rad. \textbf{b}, SrTiO$_3$ with a thickness of $78.1$\,\AA{} and a SD\,$=0.034$\,rad. \textbf{c}, Au with a thickness of $102$\,\AA{} and a SD\,$=0.035$\,rad. 
    % Both $y$ and $x$ axes represent the range of eigenvalues of both scattering and transmission matrix, respectively.
    }
    \label{fig:eig_dist}
\end{figure}

Evaluating the phases of the eigenvalues, as presented in Fig.~\ref{fig:eig_dist}, shows a very similar behavior for all three datasets when comparing the eigenvalues sorted by index. While for the scattering matrix the phase increase is perfectly linear, the phase obtained using the transmission matrix shows small deviations of up to $64$\,mrad, which appear to be independent of index and matrix size. Overall, the eigenvalues of $\mbf S$ and $\mbf{\hat S}$ are in good agreement across the analyzed range of spatial frequencies.

\subsection{Diffraction Patterns.}
\begin{figure*}[htb!]
    \centering
    \includegraphics[width=\textwidth]{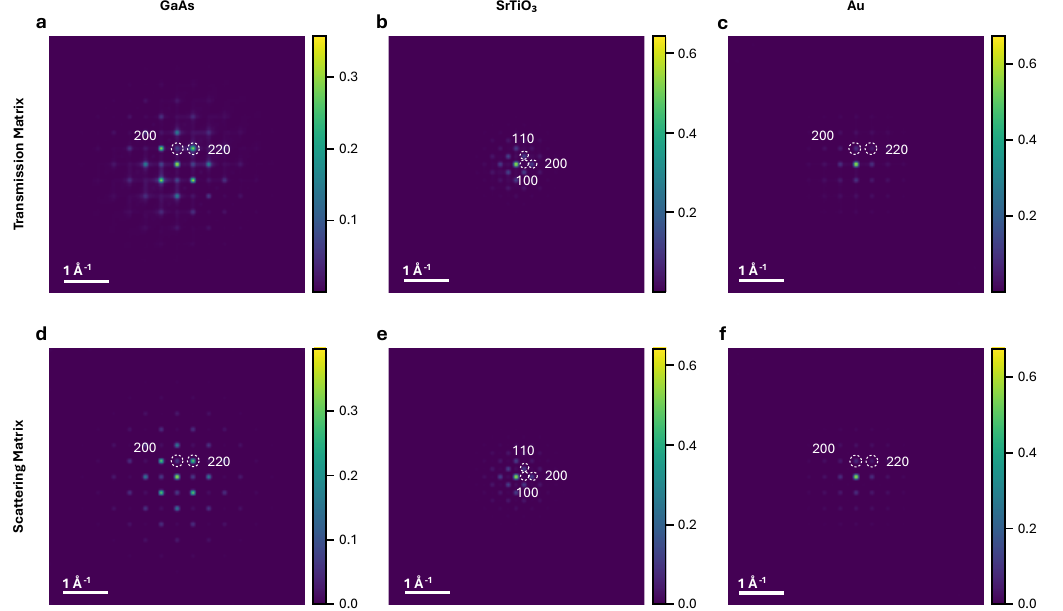}
    \caption{Plane wave diffraction patterns along the \hkl[001] zone axis calculated using the transmission matrix (top row) and the scattering matrix (bottom row). \textbf{a}, \textbf{d}, GaAs with a simulated thickness of $175.2523$\,\AA{}. \textbf{b}, \textbf{e}, SrTiO$_3$ with a simulated thickness of $126.48$\,\AA{}. \textbf{c}, \textbf{f}, Au with a simulated thickness of $121.055$\,\AA{}. The amplitude of the diffraction spots is given in arbitrary units. \ABnew{Crystalline materials are simulated with $2\times 2$ unit cells with the parameters in Table~\ref{tab:params_multislice}.}}
    % The interval of the image is from $0$ to $1$ represented by dark blue and yellow color, respectiely.
    %{need to give scale bar / pixel size, e.g. in \AA$^{-1}$. maybe plot sqrt of intensity?. (b) not convincing yet, as spots too close to each other. (c) need to check dynamic scattering in BW}
  %  } \ABnew{Could you help me with scale bar/pixel size, what is the unit? Here I already plot the amplitude or square root of intensity with dimension $45 \times 45$ pixel, what is the unit? for unit cell? The point is here only to show the equivalence between transmission and scattering matrix to produce similar Bragg peak.} {I understand. I am just worried that, if the diffraction spots dont come out properly in this figure, referees might question the results in Figure\ref{fig:bragg_scattered}}
    \label{fig:bragg_images}
\end{figure*}

The Bragg diffraction patterns of GaAs, SrTiO$_3$ and Au are evaluated as a function of specimen thickness.
% The common approach to compare both %scattering matrices 
% matrix formulations would be the evaluation of the specimen exit wave as well as %the
% its intensity.
%For the evaluation of the Bragg diffraction
To do so, the product of scattering or, respectively, transmission matrix with the illuminating plane wave is performed. The resulting amplitude of the computed diffraction patterns of all three specimens 
%in \hkl[001]-orientation 
is shown in Fig.~\ref{fig:bragg_images}.

Comparing the results obtained with the scattering matrix to those using the transmission matrix, we find a very good agreement in peak amplitude as well as pattern structure for each specimen, indicating that any inherent differences are only marginal. %can only be marginal.
\begin{figure}[htb!]
    \centering
    \includegraphics[width=\columnwidth]{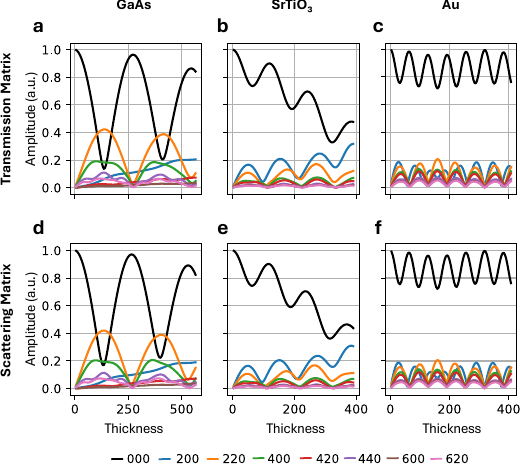}
    \caption{Bragg beam amplitude as a function of specimen thickness, calculated using the transmission matrix (top row) and the scattering matrix (bottom row). \textbf{a}, \textbf{d}, GaAs. \textbf{b}, \textbf{e}, SrTiO$_3$. \textbf{c}, \textbf{f} Au.
    %{where is the blue curve in the bottom left plot? the indices in the legend should be in curly brackets \hkl{} not \hkl[], as they are beams not directions.}
    }
    \label{fig:bragg_scattered}
\end{figure}
To go into more detail, we evaluated the propagation of several diffracted beams %by varying the
with increasing crystal thickness, 
as shown in Fig.~\ref{fig:bragg_scattered}. When propagating through the crystal, the transmitted and the diffracted beams exhibit alternating amplitudes depicting the Pendellösung effect. Our computed amplitudes using the scattering and transmission matrix are in good agreement, confirming our previous evaluation of the diffraction patterns. However, we observe slight differences, e.g. a small shift of maxima and minima in Fig.~\ref{fig:bragg_scattered}a, b, which is better visible for waves with larger oscillations.
%, as we compare two different approaches.% which is represented only one column in the matrices that are used to evaluate the scattered Bragg beam. {In the previous sentence ",which ..." part sounds strange.}
  
%BEN: (a) top: I can see additional reflections, which are not present in the bottom one (double diffraction? effect of slice thickness? not sure); (b): spots are too close together and it is hard to tell whats going on (reduce voltage for longer wavelength so spots move further apart?); main reflections need adding indices; maybe normalize/change color bar labels to only show 0 and 1?
%

\subsection{Projected Potential.}
\begin{figure}[htb!]
    \centering
    \includegraphics[width=\columnwidth]{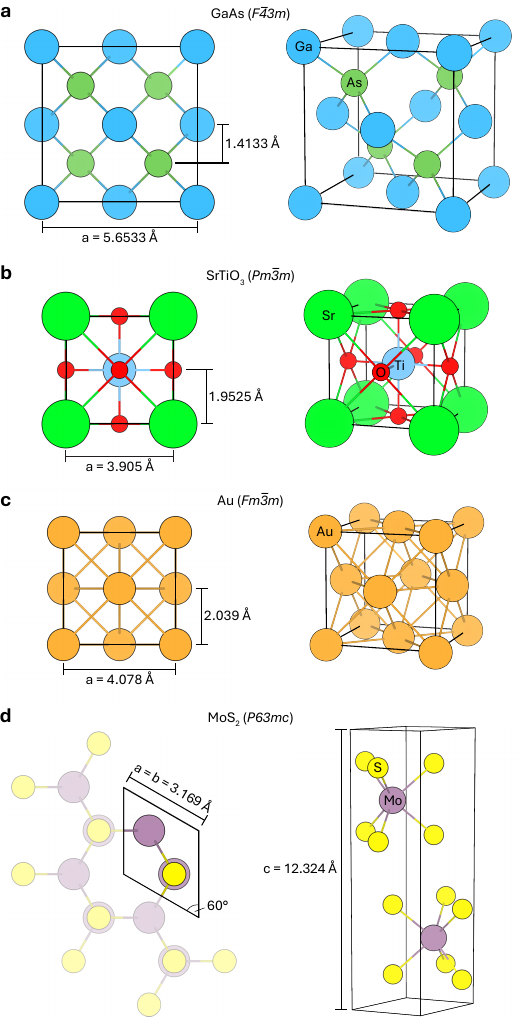}
    \caption{
     Unit cell representations in \hkl[001] projection and perspective view of the materials used to generate the scattering and the transmission matrix. \textbf{a}, gallium arsenide (GaAs) \cite{wyckoff1963interscience}. \textbf{b}, strontium titanate (SrTiO\BM{$_3$}) \cite{mitchell2000crystal}. \textbf{c}, gold (Au) \cite{wyckoff1963interscience}. \textbf{d}, molybdenum disulfide (2H-MoS$_2$) \cite{Schonfeld_a22135_MoS2} used in the 4D-STEM experiment.}
    \label{fig:materials}
\end{figure}

\begin{figure}[htb!]
    \centering
    \includegraphics[width=1\columnwidth]{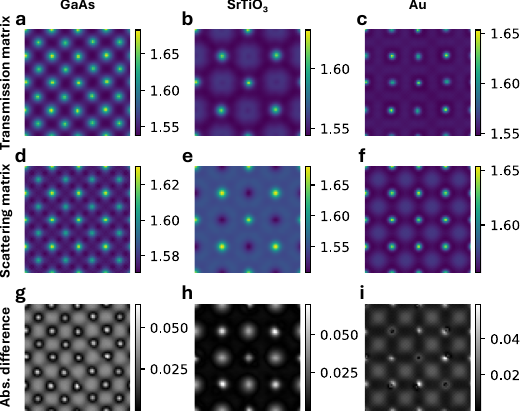}
    \caption{Projected potential of $2\times2$ unit cells in \hkl[001] zone axis of, \textbf{a}, \textbf{d}, \textbf{g}, GaAs with a thickness of $169.6$\,\AA{} (30 unit cells), \textbf{b}, \textbf{e}, \textbf{h}, SrTiO$_3$ with a thickness of $78.1$\,\AA{} (20 unit cells), and, \textbf{c}, \textbf{f}, \textbf{i}, Au with a thickness of $102$\,\AA{} (25 unit cells). The top row is calculated with the transmission matrix and the middle row with the scattering matrix. Their absolute difference is shown in the bottom row.%{Is the difference of (c) correct, it looks a little strange? could it be it has the same error as (b) before you corrected it, when the quadrants were shifted? can you also check difference (a), just in case.}}.
    %{just a suggestion: the difference could be plotted in different colour to prevent any confusion}
    }
    \label{fig:proj_potential}
\end{figure}

In the multislice method, the projected potentials are directly present in the real-space transmission matrix $\mbf{A}$, as can be seen in equation~\eqref{eq: transmission matrix}. Therefore, for the Bloch wave method, we approximate $\mbf A \approx \mbf{F}_{2D}^{-1}\mbf{S}\mbf{F}_{2D}$, up to global phase factor, from the scattering matrix $\mbf{S}$ and obtain the projected potentials.
Fig.~\ref{fig:materials}a-c shows the unit cells of the simulated crystals exhibiting their atomic structure. The same structure can be observed in the projected potentials, which are depicted in Fig.~\ref{fig:proj_potential}.
For all crystals, the calculated projected potentials retrieved using the transmission matrix in Fig.~\ref{fig:proj_potential}a-c are structurally very similar to the potentials retrieved with the scattering matrix in Fig.~\ref{fig:proj_potential}d-f. The actual values show minor deviations, as can be seen in the difference plots in Fig.~\ref{fig:proj_potential}g-i.

%, using either the scattering or the transmission matrix.

%BEN: (a) bottom: looks like the difference is not symmetric if rotated by 90degrees, why could this have happened?; what is the reason for GaAs being twice/3x as thick as the rest?; (b) middle: is it correct that at oxygen position (middle of each edge) the phase shift is lower than anywhere else in the unit cell?; should also add a scale bar here.

\begin{figure}[htb!]
    \centering
    \includegraphics[width=\columnwidth]{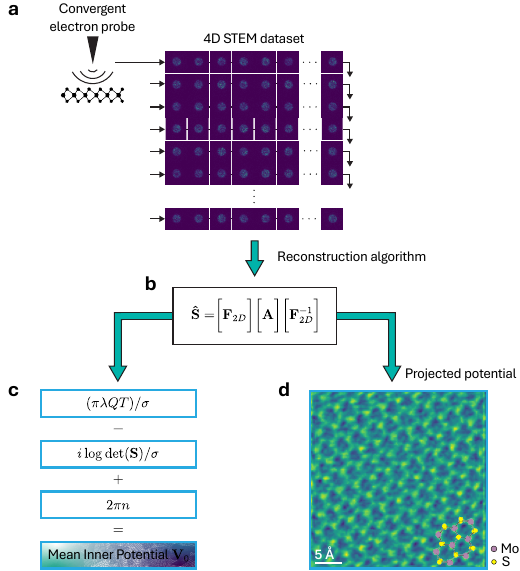}
    \caption{Workflow established for the evaluation of experimental diffraction data. \textbf{a}, 4D-STEM diffraction patterns acquired experimentally at every position of an electron transparent specimen rastered line by line using a convergent electron probe. \textbf{b}, transmission matrix reconstructed from the experimental data. \textbf{c}, estimation of the mean inner potential. \textbf{d}, projected potential of the specimen exit wave of a MoS$_2$ sample. This output is presented as is; no aberration or focus correction was applied in the reconstruction. An atomic model of MoS$_2$ is superimposed for reference.}
\label{fig:recon_mip}
\end{figure}

As an experimental example, we use the transmission matrix to calculate the projected potential from a reconstruction of a \ac{4D-STEM} dataset of MoS$_2$.
For the reconstruction, we leverage the sparse matrix decomposition algorithm \cite{bangun2022inverse}. The pipeline used for processing the experimetal data is summarized in Fig.~\ref{fig:recon_mip}. The resulting projected potential is shown in Fig.~\ref{fig:recon_mip}d, where the hexagonal atomic arrangement of MoS$_2$ can be seen. Experimental parameters are listed in Table~\ref{tab:params_stem}.

% \begin{figure}[htb!]
%     \centering
%     \includegraphics[scale=0.5]{Figures/fig_7_proj_potential_mos2_ben.pdf}
%     \caption{Projected potential {of a} MoS$_2$ {multilayer oriented close to \hkl[0001] zone axis. A schematic atomic model of MoS$_2$ is superimposed for reference. The potential was calculated from an experimental 4D-STEM dataset reconstructed by sparse matrix decomposition method from \cite{bangun2022inverse}. The output is presented as is; no aberration or focus correction was applied in the reconstruction.}}
%    %{where is thickness estimation coming from? from the reconstruction? if so, we could try verify it using the estimated thickness from PACBED. if you can adjust the focus by -1.7nm might increase resolution, if not we should mention that we didn't optimise for best resolution/image quality. also, the last column is rubbish in the dataset and should be removed. I can later add a scalebar or we mention scan width (33\,\AA). color bar needs a unit. we should include either a simulated unit cell as an inset, or the superimposed structure so that it is clear how it should look like.}}
%     %\ABnew{Here, I just included total slices and Fresnel distance. Maybe it is not correct to say thickness estimation. I dont get the colorbar unit? Could you help me on that? }
%     \label{fig:proj_potential_exp}
% \end{figure}

\subsection{Determinant and Mean Inner Potential.}
The \ac{MIP} reflects the interaction between the beam electrons and the material. It is closely linked to the inner structure of a crystal and is calculated by the average of the Coulomb potential over a unit cell, 
\begin{equation}\label{eq: MIP}
V_0 
%= \frac{1}{\Omega} \int_{\Omega} V(\mbf r) d\mbf r 
= \frac{1}{\Omega} V_{total}.    
\end{equation} 
Here, $\Omega$ is the volume of the crystal unit cell \cite{bethe1928theorie} and $V_{total}$ is the total potential. %and $\mbf r$ is the real space vectors.

The \ac{MIP} has been studied both theoretically and experimentally and was estimated for different materials in numerous works \cite{kruse2003determination,kruse2006determination,gajdardziska1993accurate,yamamoto1982surface,mizzi2021role,sun2018chemically,auslender2024mean,tull1951some,buhl1959interferenzmikroskopie,keller1961biprisma,kerschbaumer1967biprisma, goswami1982measurement, okazaki2005electronic, popescu2007increase,boureau2019quantitative}. We estimate the \ac{MIP} by employing the mathematical structure of the real space transmission matrix $\mbf{A}$, as in equation\,\eqref{eq: transmission matrix}. The determinant of $\mbf{A}$, and consequently of $\mbf{\hat S}$, is given by  
% \ABnew{The determinant of \BMnew{the }transmission matrix \BMnew{(\ref{eq: multislice scattering reciprocal1}}) is given by \BMnew{the }complex exponential function %, namely 
$$
\det (\mbf{\hat S}) \approx \det(\mbf{A}) = \exp\left(-i \left(\pi \lambda QT - \sigma V_{total}\right) \right),
$$
%where it highly depends on the parameters of wavelength of electron wave $\lambda$, total integral of reciprocal space $Q$, thickness materials $T$, interaction constant $\sigma$, and the total potential $V_{total}$.
where, $\lambda$ is the electron wavelength, $Q$ is the total integral over reciprocal space, $T$ is the specimen thickness, and $\sigma$ is the interaction constant. From this equation, we can extract the total potential %. Combining with discrete grid with pixel size parameters gives the \ac{MIP} estimate 
that, in combination with the discrete grid of pixels of known distance (step size), leads to the \ac{MIP} estimate
\begin{equation}
\label{eq:det_mip}
V_{0}^\mathrm{Det} \ABnew{= \frac{1}{\sigma N^2 c} \left( \pi \lambda   Q T  - i \log \det\left(\mbf{\hat S}\right) - 2 \pi n \right),}
\end{equation}
with $2\pi n$ for an integer $n$ resulting from the periodicity of the complex exponent. In addition, $N$ is the total number of pixels, and the unit cell parameter $c$, which is synonymous with the specimen thickness.
%\OM{Since Equation~\eqref{eq:det_mip} is prone to periodic artifacts, we also propose an alternative technique for estimating MIP based on phase unwrapping $\mbf{V}_0^\mathrm{Unwrap}$.} \ABnew{The estimation is directly performed by calculating the total potentials of simulated or experimental slices from reconstruction method, i.e., phase of the object $\mbf{O}_m$ for each slice $m$ with unwrapping technique.}
Since equation~\eqref{eq:det_mip} is prone to periodic artifacts, we follow an alternative route based on phase unwrapping for estimating the \ac{MIP} $\mbf{V}_0^\mathrm{Unwrap}$. Our estimation is directly performed by calculating the total potential of each slice, i.e. the phase of the object $\mbf{O}_m$ of each slice $m$, using the unwrapping technique.

\begin{table*}[hbt!]
  \caption{Estimates of the \acl{MIP} $\mbf{V}_0^\mathrm{Det}$ from the determinant of the transmission matrix using equation\,\eqref{eq:det_mip} \ABnew{ and \acl{MIP} $\mbf{V}_0^\mathrm{Unwrap}$ by applying phase unwrap of each estimated slice before calculating MIP} in comparison to literature values from calculated ($\mbf{V}_0^\mathrm{DFT}$) and experimental ($\mbf{V}_0^\mathrm{Exp}$) data. The value of $n$, the periodicity from the complex exponent, is chosen such that $\mbf{V}_0^\mathrm{Det}$ is closer to $\mbf{V}_0^\mathrm{Unwrap}$. Values for $\mbf{V}_0^\mathrm{Det}$ and $\mbf{V}_0^\mathrm{Unwrap}$ are labeled with (sim.) for simulated and (exp.) for experimental data.}%\BMnew{BEN: what is meant with factor C, it isn't anywhere mentioned before, and why is it important to mention it here? can it go in the text?} .\BMnew{BEN: need to mention MIP after unwrapping. Also: need to check how supercell of STO is terminated (is the interface SrO/vac or TiO2/vac), then we can compare our result to the respective literature values}from both simulated and experimental data
\label{tab:mip}
  \small
  \centering
  \begin{tabular}{lllll}
    \toprule
     \bfseries Material & $\mbf{V}_0^\mathrm{DFT}$ (V)& $\mbf{V}_0^\mathrm{Exp}$ (V)&   $\mbf{V}_0^\mathrm{Det}$ (V) & \OM{$\mbf{V}_0^\mathrm{Unwrap}$ (V)}
     \\
     \midrule
      GaAs  & $14.19$  \cite{kruse2006determination} & $14.24\pm 0.08$ \cite{kruse2003determination,kruse2006determination} & $14.33 + 0.403  \, n, \ n \in \mbb{Z}$ {(sim.)}& $14.89$ (sim.) \\
       &
       & $14.53 \pm 0.17$ \cite{gajdardziska1993accurate, kruse2006determination}, $13.2$ \cite{yamamoto1982surface} \\
      SrTiO$_3$  & SrO $15.2$ \cite{mizzi2021role}, TiO$_2$ $17.7$ \cite{mizzi2021role} & SrO $13.3$\cite{sun2018chemically}, TiO$_2$ $14.6$ \cite{sun2018chemically} & SrO \OM{
      $17.92 +0.58\, n, \ n \in \mbb{Z}$ {(sim.)}}
      %$18.5 +0.58\, z, \ z \in \mbb{Z}$ {(sim.)}
      & SrO $18.12$ (sim.) \\
      Au  &  $28.40$ \cite{auslender2024mean} & $30.2$\cite{tull1951some}, $21.2$ \cite{buhl1959interferenzmikroskopie}, $22-27$ \cite{keller1961biprisma} & $28.54 + 0.558\, n, \ n \in \mbb{Z}$ {(sim.)}& $28.17$ (sim.)\\ 
      & & $16.8$ \cite{kerschbaumer1967biprisma},$21.4$ \cite{goswami1982measurement},
      $25$ \cite{okazaki2005electronic}\\
      & & $32.2$ \cite{popescu2007increase}
      \\
      MoS$_2$ & $10.6$ \cite{boureau2019quantitative} & $10.4$ \cite{boureau2019quantitative} &  $10.41 + 0.011\, n, \ n \in \mbb{Z}$ {(exp.)}&  $10.74$ (exp.)\\
    \bottomrule
  \end{tabular}
\end{table*} 

% \paragraph{Mean Inner Potential} 
The estimation of the \ac{MIP} $\mbf{V}_0$ for simulated and experimental datasets is given in Table~\ref{tab:mip}. For comparison, results from density functional theory (DFT) calculations as well as experimental data reported in other studies is provided. %In general, by using the determinant up to modulo $2\pi n$ for integer $n$, we are able to estimate the \ac{MIP} of the materials considered in our study for both simulated and experimental data in very good agreement with previously reported values.
%By considering the determinant up to modulo $2\pi n$ for integers of $n$, we were able to estimate the \ac{MIP} in very good agreement with previously reported values, using both the simulated and experimental diffraction data evaluated in our study.
By considering the determinant up to modulo $2\pi n$ for integers of $n$, we were able to estimate the \ac{MIP} using both simulated and experimental diffraction data in very good agreement with previously reported values.

\subsection{Computational Complexity.}
The apparent strength of the multislice over the Bloch wave method is the computational complexity. To compute the $N^2 \times N^2$ scattering matrix via equation~\eqref{eq:scatmat_main}, eigenvalue decomposition 
%of the structure matrix 
has to be performed, requiring $\mcl{O}\left( N^6 \right)$ operations. For the transmission matrix, we perform the sequential matrix product of Fresnel matrices $\mbf{G}_m$ and diagonal potential matrices $\mbf{O}_m$. The products with $\mbf{O}_m$ and $\mbf{G}_m$ require $\mcl{O}\left(N^4 \right)$ and, respectively, $\mcl{O}\left(N^4 \log N \right)$ operations. In the first case, we use that $\mbf{O}_m$ is diagonal, and in the second, we employ the fast Fourier transform. Repeating these steps for $M$ slices %gives the
leads to a total complexity of $\mcl{O}\left(M N^4 \log N \right)$ operations.

\section{Discussion}
In this study, we discussed a new perspective for analyzing and understanding the relationship between the multislice and Bloch wave methods.
%In conclusion, t
Our presented reformulation of the multislice method into a matrix framework, which we call transmission matrix, %is 
not only %possible
enables %to 
leveraging eigenstructure analysis to measure the similarity to the scattering matrix but also %to
allows the estimation of the \ac{MIP} %of crystalline materials 
from diffraction data.

By expressing the multislice method as the transmission matrix, we have demonstrated that its structure parallels that of the scattering matrix. %Theoretically, w
Firstly, we demonstrate mathematically that the eigenvalues of both matrices %should be 
are equal up to an additive multiple of $2\pi$. %$\pm 2\pi n$ where $n$ is the natural number.  
Secondly, we prove that the eigenvectors are interconnected through the two-dimensional Fourier matrix, with the eigenvector of the transmission matrix representing the Bloch wave itself in %the
real space. %This

Our matrix-based approach %also 
allows %us to 
a consistent approximation of the projected potential% of the material% from the scattering matrix
. Moreover, our numerical evaluations of plane-wave diffraction, eigenvalues, as well as projected and \ac{MIP} support these theoretical results, confirming the robustness and accuracy of the theoretical derivation of eigenstructures for both the transmission and scattering matrix. %ces.
%This approach discusses a new perspective for analyzing and understanding the relationship between multislice and Bloch wave methods. {adjusted and moved up}

Furthermore, the structure of the transmission matrix enables the utilization of the determinant, i.e., the product of all eigenvalues, to estimate the \ac{MIP} of crystalline materials% for both simulated and experimental data
, making it a convenient tool for the evaluation of practical experiments, particularly due to its lower computational demands relative to the scattering matrix of the Bloch wave approach.

%{Should we include a brief comparison of the computational cost, so that we can state this as an additional advantage? No figure needed, just timing both methods should be sufficient, and mention the CPU/RAM of the system used.}
%\ABnew{I think adding  computation highly depends on the machine. We have added the computational complexity section.}

%{Styling of the references is not consistent. Sometimes it shows full first names, sometimes just the initials. In [2] and [19] there are undescores after the initials.}
%\ABnew{I try to fix all, but it seems i could not find all names}

\BMnew{To put our results in Table~\ref{tab:mip} into perspective, our multislice simulations were conducted using ideal crystals without considering surface relaxations of the terminating layer at the specimen/vacuum interface, that in reality would be expected. Our estimated \ac{MIP}s therefore do not account for any specific surface contributions. It is thus noteworthy that, nevertheless, our results compare well to experimentally determined values of real-world samples ($\mbf{V}_0^\mathrm{Exp}$), and to results of DFT calculations of relaxed structures ($\mbf{V}_0^\mathrm{DFT}$) previously reported. Only the estimation of the \ac{MIP} of SrTiO$_3$ remains a challenge which we attribute to its strong ionic character.
}
 
%\BMnew{Due to the surface-sensitivity of the \ac{MIP}, we attribute the deviation of our estimated \ac{MIP} of MoS$_2$ after phase unwrapping to potential surface contamination of the specimen during the \ac{4D-STEM} experiment.}

\BMnew{We estimate that our presented method for determining the \ac{MIP} to perform equally well or better with any experimental \ac{4D-STEM} data as well as with simulated \ac{4D-STEM} data from relaxed structures.}
\section{Methods}\label{sec:methods}
%We give a brief introduction to the 3D representation of the transmission function %in
%of crystalline materials% will be presented
%, which is commonly used in the \ac{TEM} %transmission electron microscopy 
%nomenclature.
\subsection{Notation.}
Vectors are written in bold small-cap letters $\mathbf{x} \in \C^L$ and matrices are written in bold big-cap letter{s} $\mathbf{A} \in \C^{K \times L}$ for a complex field $\C$ and $\R$ for a real field. Matrices can also be written as its element 
$\mathbf{A} = \left(a_{k\ell}\right)$, where $k \in [K], \ell \in [L]$. The set of integers is written as $[N] := \{1,2,\hdots,N\}$.
%and calligraphic letters are used to define functions $\mathcal{A} : \C \rightarrow \C$. \OM{Specifically, we denote the discrete two{-}dimensional Fourier transform by $\mathcal{F}$.}
For both matrices and vectors the notation $\circ$ is used to represent the element-wise or Hadamard product. The $\mbf{A}^H$ is used to represent conjugate transpose. A matrix is called unitary if \OM{$\mbf{A}^H\mbf{A} = \mbf{A}\mbf{A}^H = \mbf{I}$}, where $\mbf{I}$ is the identity matrix. For a matrix $\mbf A \in \C^{K \times L}$, the Frobenius norm is given by $$\norm{\mbf A}_F = \left(\sum_{k = 1}^K \sum_{l=1}^L \card{a_{kl}}^2\right)^{1/2} = \left( \trace \left( \mbf A^H\mbf A\right) \right)^{1/2}.$$

\subsection{Scattering Matrix.}
%\AB{Need to discuss: } {RANDOM THOUGHT: SHOULD WE USE ABREVIATION SM FOR SCATTERING MATRIX?} {S-matrix is commonly used}

The derivation of the scattering matrix in \ac{TEM} %transmission electron microscopy 
is tied to the theory of dynamical electron scattering in crystals. 
% \OM{Starting from the non-relativistic stationary Schrödinger equation, with relativistic corrections made by Dirac, and by neglecting the spin, the Klein-Gordon equation can be used to describe the scattering of relativistic electrons in \ac{TEM}}:%transmission electron microscopy:}  % {} as follows,
We start with Klein-Gordon equation to describe the relativistic scattering of electrons in the crystal
\begin{equation}\label{eq: Schroedinger}
     \Delta \Psi\left(\mbf{r}\right)  + 4\pi^2k_0^2\Psi\left(\mbf{r}\right)  = - \frac{8\pi^2me}{h^2}V\left(\mbf{r}\right) \Psi\left(\mbf{r}\right) ,
\end{equation}
where $V$ is the electrostatic potential at three-dimensional real space coordinates $\mbf r = (x, y, z)^T$, $m$ is the relativistic electron mass, $h$ is the Planck constant, %and
$e$ is the elementary charge%. %{ISN'T WAVE NUMBER USUALLY DENOTED BY A SMALL LETTER?}
%Additionally, the 
, and $k_0$ is the relativistic wave number.
Intuitively, the goal is to find proper wave functions $\Psi$ that fulfill the Klein-Gordon equation. %, \OM{specifically to quantify the scattering processes in the crystal.}

As discussed in \cite[eq. 5.43]{de2003introduction}, the strategy to solve the partial differential equation~\eqref{eq: Schroedinger} is by involving the expansion in the Fourier space, also called reciprocal space. Firstly, the potential is expanded into its Fourier series,
$$
V\left( \mbf r \right) = \sum_{\mbf h} V_{\mbf h} \exp\left(i 2\pi \mbf h \cdot \mbf r\right),
$$
where $ V_{\mbf h}$ is the lattice potential in reciprocal space. Secondly, the Bloch wave is incorporated, given as
$$
\Psi\left(\mbf{r}\right) = c\left(\mbf{r}\right)\exp\left(i2\pi {\mbf{k}}\cdot{\mbf{r}}\right) = \sum_{\mbf g} c_{\mbf g} \exp\left(i 2\pi {\left(\mbf k + \mbf g\right)} \cdot {\mbf r}\right),
$$
where $\mbf k, \mbf g \in \R^3$ are wave %vector 
and reciprocal space vectors, respectively. Variable $c_{\mbf g}$ is the Bloch wave coefficient %s 
in reciprocal space. %R
By rearranging this equation and writing the wave functions and potential in reciprocal space coordinates, we obtain the eigenvalue equation, the solution of which is the scattering matrix 
$$
\mbf{S} = \exp\left(i 2\pi T \mbf{B}\right).
$$
For a complete derivation, we refer the interested readers to the literature, for instance \cite{metherell1976diffraction, humphreys1979scattering,de2003introduction}.
Here $\mbf{B} \in \C^{N^2 \times N^2}$ is the structure matrix representing the electrostatic potential of the underlying crystal given by
\begin{equation*}
\label{eq:structure_matrix}
\begin{pmatrix}
  & \vdots & \vdots & \vdots  & \vdots & \vdots &  \\
\hdots &s_{\mbf h} & V_{\mbf h - \mbf g}   & V_{\mbf h}  &  V_{\mbf h + \mbf g} & V_{\mbf 2h} & \hdots \\
\hdots &V_{\mbf g - \mbf h} & s_{\mbf g} & V_{\mbf g} & V_{2\mbf g}&  V_{\mbf h + \mbf g} &\hdots\\
\hdots &V_{ - \mbf h} & V_{- \mbf g} & s_{\bs 0 } & V_{\mbf g}  &  V_{\mbf h}&{\hdots}\\
\hdots & V_{-\mbf g - \mbf h} & V_{-2\mbf g} & V_{-\mbf g} & s_{-\mbf g} &  V_{-\mbf g + \mbf h}&{\hdots}\\
\hdots &V_{-2\mbf h} & V_{-\mbf h-\mbf g} & V_{-\mbf h} & V_{-\mbf h + \mbf g} &  s_{-\mbf h} &{\hdots}\\
 &\vdots & \vdots & \vdots  & \vdots &  {\vdots}\\
\end{pmatrix},
\end{equation*}
where the diagonal and off-diagonal elements represent the excitation error and, respectively, potential in reciprocal space, and $\mbf h \in \R^3$ and $\mbf g \in \R^3$ are the reciprocal lattice vectors. This relation between the crystal and the scattering matrix is illustrated in Fig.~\ref{fig:blochwave_vs_multislice}a, c.
Due to the nature of matrix exponentiation, both $\mbf S$ and $\mbf B$ share eigenvectors $\mbf C$, 
\begin{equation*}
\label{eq:eigvect}
\mbf C = \begin{pmatrix}
\vdots & \vdots & \vdots & \vdots  & \vdots &  \\
c_{\mbf h}^1 & c_{\mbf h}^2 & \hdots & c_{\mbf h}^i &  \hdots\\
c_{\mbf g}^1 & c_{\mbf g}^2 & \hdots & c_{\mbf g}^i &  \hdots\\
c_{\mbf 0}^1 & c_{\mbf 0}^2 & \hdots & c_{\mbf 0}^i &  \hdots\\
c_{-\mbf g}^1 & c_{-\mbf g}^2 & \hdots & c_{-\mbf g}^i &  \hdots\\
c_{-\mbf h}^1 & c_{-\mbf h}^2 & \hdots & c_{-\mbf h}^i &  \hdots\\
\vdots & \vdots & \vdots  & \vdots & \vdots \\
\end{pmatrix}	\in \C^{N^2 \times N^2}.
\end{equation*}
Furthermore, the eigenvalues $\Lambda$ of $\mbf S$ are given as
\[ 
\left(\exp{\left(2\pi i \gamma_1 T\right)},\exp{\left(2\pi i \gamma_2 T\right)},\hdots, \exp{\left(2\pi i \gamma_{N^2} T\right)}\right).
\]
where $\gamma_j$, $j \in [N^2]$ are the eigenvalues of $\mbf B$. Combined, this yields equation~\eqref{eq:scatmat_main}.

\subsection{Transmission Matrix.}
% Apart from the Bloch wave-derived scattering matrix, the multislice method is frequently used to model the interaction between an incident electron wave and the crystal potential.

Instead of directly deriving an analytical solution as is done with the Bloch wave approach, the transmission function can be constructed using multiple infinitely thin (sub-)specimens, i.e., slices, separated by free space. This is the approach we take using the multislice method.

More precisely, let $V$ denote the potential energy of the three-dimensional crystal. The projection of $V$ onto the $x$-$y$ plane at $z_m$ is expressed by $V_m(x,y) = \int_{z_m}^{z_{m+1}} V(x,y,z) \mathrm{d}z$. Then, the projected potentials $V_m$ are arranged as diagonal matrices $\mbf{O}_m \in \C^{N^2 \times N^2}$ of equation~\eqref{eq: transmission matrix}, each representing the $m$-th slice. Its entries are given by 
\begin{equation}
\label{eq:potential}
\left(\mbf O_m\right)_{x + N(y-1), x +N(y-1)} = \exp\left(i\sigma V_m(x,y) \right) \,\text{for}\,x,y \in [N],
\end{equation} 
with the interaction constant $\sigma$ for relativistic correction. For modeling the scattering process, we also account for the separation of the slices $\mbf O_m$ and $\mbf O_{m+1}$ by distance $\Delta_m = z_{m+1}-z_{m}$ as depicted in Fig.~\ref{fig:blochwave_vs_multislice}b. This results in a free space propagation described by the Fresnel matrix $\mbf G_m$,
\begin{equation}
\label{eq:fresnel_matrix}
\mathbf{G}_m \defeq \mathbf{F}^{-1}_{2D}\mbf{D}_m\mathbf{F}_{2D} \in \C^{N^2 \times N^2}, \ m \in [M],
\end{equation}
where $\mbf{D}_m \in \C^{N^2 \times N^2}$ is a diagonal matrix. Its elements are
\begin{equation}
\left(\mbf{D}_m\right)_{x + N(y-1), x +N(y-1)}
 \defeq \exp\left(-\pi i\Delta_m \lambda Q_{xy}\right)\,\text{for}\, x,y\in [N],
\label{eq:fresnel_element}
\end{equation}
with $Q_{xy} = \left( q_y^2 + q_x^2 \right) + 2\left( q_x \frac{\sin \theta_x}{\lambda} + q_y \frac{\sin \theta_y}{\lambda} \right)$ defined by the spatial frequencies $q_x,q_y$ in reciprocal space and two-dimensional tilt angles $\theta_x, \theta_y$ describing the direction of the incoming wave. 

Then, the scattering of the incident beam within the crystal in the multislice model is obtained by consecutive multiplication with slice $\mbf O_m$ and Fresnel propagators $\mbf G_m$. Combining all multiplications together yields the real space transmission matrix $\mbf A$ and its reciprocal space counterpart $\mbf{\hat S}$ as in equations~\eqref{eq: transmission matrix} and \eqref{eq: multislice scattering reciprocal1}. For the detailed derivation, see \cite{bangun2022inverse}.
%\OM{An illustration visualizing the multislice method is shown in Figure \ref{fig:blochwave_vs_multislice}{(b)}.}

As highlighted in \cite{cowley1957scattering,de2003introduction}, the multislice method serves as an alternative to the Bloch wave formulation, making the transmission matrix $\mbf{\hat S}$ 
%which we have derived from the multislice method 
an approximate solution to the Schrödinger equation~\eqref{eq: Schroedinger}. The main advantage that this reformulation offers is that the atomic structure of a crystalline object can now be represented independently by the transmission matrix $\mbf{\hat S}$, without being influenced by the electron probe, which is not the case in the classical multislice model.

\subsection{Characterizing Similarity.}

In the following, we compare the transmission matrix $\mbf{\hat S}$ in equation~\eqref{eq: multislice scattering reciprocal1} and scattering matrix $\mbf{S}$ in equation~\eqref{eq:scatmat_main} in terms of their eigenstructures. 
%\A{We {will} also draw a connection to estimate {the} \acl{MIP} from {the }transmission matrix.}

%$ = \mbf{C} \bs \Lambda \mbf{C}^{-1}$.
%\AB{Need to discuss: }  {PARAGRAPH COMMENTED OUT.}
% We construct  matrices from both multislice and Bloch wave method for a given thick specimen and provide analysis of the structure  in terms of their eigenvalues  and eigenvectors. For scattering matrix from Bloch wave formulation $\mbf{S}$, the eigenvalues and eigenvectors are well-defined and presented in \eqref{eq:scatmat}. However, for the matrix $\mbf{\hat S}$  constructed from multislice method, the structure of eigenvalues and eigenvectors is unknown.

As $\mbf{S}$ is defined via eigenvalue decomposition in equation~\eqref{eq:scatmat_main}, we first establish the eigendecomposition for $\mbf{\hat S}$. 
% First, we investigate the eigenvalue decomposition of $\mbf{\hat S} = \mbf{F}_{2D} \mbf{A} \mbf{F}_{2D}^{-1}$. 
By construction, we have
\[
\mbf{\hat S} = \mbf{F}_{2D} \prod_{m = 1}^{M}\mathbf{G}_{m}\mbf{O}_m \mbf{F}_{2D}^{-1}
= \mbf{F}_{2D} \prod_{m = 1}^{M} \mbf{F}_{2D}^{-1} \mbf{D}_m\mbf{F}_{2D}\mbf{O}_m \mbf{F}_{2D}^{-1},
\]
where matrices $\mbf{D}_m$ and $\mbf{O}_m$ are both diagonal with unit modulus entries, and are therefore unitary. Fourier transforms are also unitary and, consequently, $\mbf{\hat S}$ is unitary %as
being the product of unitary matrices. Hence, this also implies that $\mbf{\hat S}$ is normal matrix, i.e., it satisfies $\mbf{\hat S}^H \mbf{\hat S} = \mbf{\hat S} \mbf{\hat S}^H = \mbf{I}$. As a result, $\mbf{\hat S}$ admits eigenvalue decomposition $\mbf{\hat S} = \mbf{W} \Sigma \mbf{W}^{-1}$ with a unitary matrix $\mbf{W}$, %whose
which columns are the eigenvectors of $\mbf{\hat S}$, and with a diagonal matrix $\Sigma$ with eigenvalues of $\mbf{\hat S}$ on its diagonal. Since $\mbf{\hat S}$ is unitary, the eigenvalues are of unit modulus and can be written in terms of a complex exponential, $\lambda_k\left(\mbf{\hat S}\right) = \exp\left(i2\pi\theta_k \right)$, where $\theta_k \in [0,1]$ for $k \in [N^2]$.

To quantify any differences, the scattering and the transmission matrix are compared in terms of the Frobenius norm. This norm acts as a distance metric, enabling the assessment of the accuracy and consistency between both matrices. 
{For simplicity, we assume that the eigenvalues and eigenvectors of $\mbf S$ and $\mbf{\hat S}$ are sorted in the order of increasing angles $\gamma_k$ and $\theta_k$. 
Then, using triangle inequality, we split the error into the eigenvector and eigenvalue errors as
\begin{align*}
& \norm{\mbf{S} - \mbf{\hat{S}}}_F
= \norm{\mbf{C}\Lambda\mbf{C}^{-1} - \mbf{W} \Sigma \mbf{W}^{-1}}_F \\
& \quad \le \norm{\mbf{C}\Lambda\mbf{C}^{-1} - \mbf{W} \Lambda \mbf{W}^{-1}}_F
+ \norm{\mbf{W}\Lambda\mbf{W}^{-1} - \mbf{W} \Sigma \mbf{W}^{-1}}_F.
\end{align*}
We further transform the first term as
\begin{align*}
& \norm{\mbf{C}\Lambda\mbf{C}^{-1} - \mbf{W} \Lambda \mbf{W}^{-1}}_F \\
& \quad \le \norm{\mbf{C}\Lambda\mbf{C}^{-1} - \mbf{W} \Lambda \mbf{C}^{-1}}_F 
+ \norm{\mbf{W}\Lambda\mbf{C}^{-1} - \mbf{W} \Lambda \mbf{W}^{-1}}_F
\\
& \quad = \norm{ (\mbf{C} - \mbf{W}) \Lambda\mbf{C}^{-1} }_F 
+ \norm{\mbf{W}\Lambda (\mbf{C}^{-1} - \mbf{W}^{-1} )}_F \\
& \quad \le \norm{ \mbf{C} - \mbf{W} }_F \norm{ \Lambda \mbf{C}^{-1} }_2 
+ \norm{ \mbf{C}^{-1} - \mbf{W}^{-1} }_F\norm{ \mbf{W} \Lambda}_2,
\end{align*}
where in the second line we used the triangle inequality again and in the last line we used that Frobenius norm of the product is bounded by the product of Frobenius and spectral norms. The spectral norm is the magnitude of the largest eigenvalue and for unitary $\Lambda \mbf{C}^{-1}$ and $\mbf{W} \Lambda$ it is equal to one. Moreover, since both $\mbf C$ and $\mbf W$ are unitary, we get $\mbf C^{-1} = \mbf C^H$, $\mbf W^{-1} = \mbf W^H$ and  
\[
\norm{ \mbf{C}^{-1} - \mbf{W}^{-1} }_F 
= \norm{ (\mbf{C} - \mbf{W})^{H} }_F
= \norm{ \mbf{C} - \mbf{W} }_F.
\]
Consequently, the upper bound for the first term simplifies to
\[
\norm{\mbf{C}\Lambda\mbf{C}^{-1} - \mbf{W} \Lambda \mbf{W}^{-1}}_F \le 2 \norm{ \mbf{C} - \mbf{W} }_F,
\]
which is the error between the eigenvector matrices. Similarly, we bound the second term as the error on the eigenvalues,
\begin{align*}
& \norm{\mbf{W}\Lambda\mbf{W}^{-1} - \mbf{W} \Sigma \mbf{W}^{-1}}_F
 = \norm{\mbf{W} (\Lambda - \Sigma) \mbf{W}^{-1}}_F \\
& \quad \le \norm{\mbf{W}}_2 \norm{\Lambda - \Sigma }_F \norm{\mbf{W}^{-1}}_2 = \norm{\Lambda - \Sigma }_F.
\end{align*}
Combined, it yields
\[
\norm{\mbf{S} - \mbf{\hat{S}}}_F \le 2 \norm{ \mbf{C} - \mbf{W} }_F + \norm{\Lambda - \Sigma }_F.
\]
We can further expand the eigenvalue error in terms of angles,
\begin{align}
& \norm{\Lambda - \Sigma }_F^2  
=  2 N^2  -  \text{trace}\left( \Lambda^H  \Sigma\right) - \text{trace}\left( \Lambda  \Sigma^H\right), \nonumber \\
& \quad \,\stackrel{\text{(c)}}{=} 2N^2 -  \sum_{k=1}^{N^2}[ e^{\left(i2\pi\left(\gamma_kT - \theta_k\right)\right)} + e^{\left(-i2\pi\left(\gamma_kT - \theta_k \right)\right)}] \nonumber \\
& \quad \,\stackrel{\text{(d)}}{=} 2N^2 - 2\sum_{k=1}^{N^2}\cos\left(2\pi \left(\gamma_kT - \theta_k\right)\right){.}
\label{eq:equivalence}
\end{align}

The equality $(c)$ is derived from the fact that the eigenvalue of matrix $\mbf S$ is $\exp(i2\pi\gamma_k T)$ for $k \in [N^2]$ and that matrix $\mbf{A}$ is a unitary matrix %that has 
with modulus eigenvalues of the %in 
form $\exp(i2\pi\theta_k)$ for $k \in [N^2]$.
% \begin{comment}
% \ABnew{In the following we should remove since we move into results}
% {Therefore, the equivalence between the two matrices in terms of the Frobenius norm holds as long as the following conditions are satisfied:}
% \begin{itemize}
%     \item Eigenvectors $\mbf C$ of scattering matrix $\mbf S$ and eigenvector{s} $\mbf W$ of matrix $\mbf A$ are related in terms of  a two-dimensional Fourier matrix, namely $\mbf{C} = \mbf{F}_{2D} \mbf{W}$.
%     \item The angles of {both the eigenvalue matrix} %matrix eigenvalues 
%     $\Lambda${, obtained} from scattering matrix $\mbf S${,} and {the eigenvalue matrix} %matrix eigenvalues 
%     $\mbf{V}${, obtained} from matrix $\mbf A${,} differ by up to modulo $2\pi n$ for integer{s} $n = \gamma_k T - \theta_k$. 
% \end{itemize} 
% \begin{remark}
%  {%We can also approximate the m
%  {M}atrix $\mbf{A}$ {can also be approximated} by applying two-dimensional Fourier transform and inverse Fourier transform to the scattering matrix, i.e., $\mbf A \approx \mbf{F}_{2D}^{-1}\mbf{S}\mbf{F}_{2D}$. This is true, because the eigenvector of the scattering matrix $\mbf S$ is the Fourier coefficient %s 
%  of {the} Bloch wave, as discussed in \eqref{eq:eigvect}. Consequently, the same total projected potential can be obtained from both matrix $\mbf A$  and  $\mbf{F}_{2D}^{-1}\mbf{S}\mbf{F}_{2D}$. We will show numerically in Section  \ref{sec:numerics}, that both matrices produce a very similar projected potential.}  
% \end{remark}}
% \end{comment}
%
\subsection{Determinant and Mean Inner Potential.}
Apart from characterizing similarity, we can also estimate the \acf{MIP} using the transmission and scattering matrices. As defined in equation~\eqref{eq: MIP}, it requires the total potential $V_{total}$ that is estimated using the determinant of $\mbf{\hat S}$. More precisely, we compute $\det(\mbf{\hat S})$ as 
% from the eigenstructure in terms of the determinant. 
% The determinant of the scattering matrix $\mbf S$ is well-known and can be derived as 
% \[
% \det(\mbf S) = \prod_{j = 1}^{N^2} \exp\left(2\pi i \gamma_j T\right) = \exp\left(2\pi i T \gamma_{total}\right),
% \]
% where $T$ is the thickness parameter and $\gamma_{total} = \sum_{j=1}^{N^2} \gamma_j$. 
% For the transmission matrix $\mbf{\hat S}$, the determinant can be derived as follows determinant as follows
\begin{equation}\label{eq:det product}
    \begin{aligned}
    \det\left(\mbf{\hat S}\right)
    &=\det\left(\mbf{F}_{2D}\right)\det\left(\mbf{A}\right) \det\left(\mbf{F}_{2D}^{-1}\right) \\&= \det \left( \prod_{m = 1}^{M}\mathbf{G}_{m}\mbf{O}_m \right)\\&= \det \left(\mathbf{G}_{1}\mbf{O}_1 \right)\det \left(\mathbf{G}_{2}\mbf{O}_2 \right) \dots \det \left(\mathbf{G}_{M}\mbf{O}_M \right).
    \end{aligned}
    \end{equation}
    The equality holds because {$\det(\mbf{F}_{2D}^{-1}) = 1/\det(\mbf{F}_{2D})$} and the determinant of a matrix product is equal to the product of the determinants of the individual factors. Usign that the determinant of  diagonal matrix is the product of the elements, the determinant of products $\mbf{G}_m \mbf{O}_m$
    \begin{equation*}\label{eq: det slice}
    \begin{aligned}
        \det\left( \mbf{G}_m \mbf{O}_m \right)
        & \stackrel{\eqref{eq:fresnel_matrix}}{=} \det\left(\mbf{F}_{2D}^{-1} \mbf{D}_m \mbf{F}_{2D}\right) \det\left( \mbf{O}_m\right)\\
        & = \det\left( \mbf{G}_m\right) \det\left( \mbf{O}_m\right) \\
        & \stackrel{\eqref{eq:potential}, \eqref{eq:fresnel_element}}{=} \prod_{x,y = 1}^N\exp\left(-\pi i\Delta_m \lambda Q_{xy}\right) \exp\left(i\sigma V_m(x,y) \right)\\ 
        & = \exp\left(-i\left(\pi \Delta_m \lambda Q - \sigma V_m \right) \right) \\
        \end{aligned}
    \end{equation*}
    where we used notation 
    \[
    Q = \sum_{x=1}^N \sum_{y=1}^N Q_{xy} 
    \quad \text{and} \quad 
    V_m = \sum_{x=1}^N \sum_{y=1}^N V_m(x,y).
    \]
    Returning to equation~\eqref{eq:det product}, the logarithm of the determinant of $\mbf{\hat S}$ is
    % \begin{equation}
    % \begin{aligned}
    %  \label{eq:det_phase}
    %     \det\left(\mbf{\hat S}\right)
    %     &\stackrel{\text{(e)}}{=}  \prod_{m=1}^M \exp\left(-i\left(\pi \Delta_m \lambda   Q - \sigma  V_m \right) \right) \\
    %     &\stackrel{\text{(f)}}{=}\exp\left(-i\left(\pi \lambda   Q\sum_{m=1}^M\Delta_m  - \sigma  \sum_{m=1}^MV_m \right) \right) \\
    %     &\stackrel{\text{(g)}}{=}\exp\left(-i\left(\pi \lambda   Q T - \sigma  V_{total} \right) \right).
    %     \end{aligned}
    % \end{equation}
    \begin{equation}
    \begin{aligned}
     \label{eq:det_phase}
        \log \det\left(\mbf{\hat S}\right)
        & = -i \pi \lambda Q\sum_{m=1}^M\Delta_m  + i \sigma \sum_{m=1}^M V_m + 2 \pi i n  \\
        & = -i \pi \lambda   Q T + i \sigma \sum_{m=1}^M V_m + 2 \pi i n .
        \end{aligned}
    \end{equation}
    where $T = \sum_{m=1}^M \Delta_m$ is the total thickness of the crystal and $2\pi i n$ with integer $n$ results from periodicity of complex logarithm. 

    \OM{The sum of potentials is linked to the total potential
    \begin{equation*}
    \begin{aligned}
    V_{total} & =  \int_x \int_y  \int_z V(x,y,z) \mrm{d}_z \mrm{d}_y \mrm{d}_x
    \end{aligned}
    \end{equation*}
    by the Riemman sum approximation. Following equation C.20 in Kirkland, 1998 \cite{kirkland1998advanced}, the inner integral is precisely $\int_z V(x,y,z) \mrm{d}_z = \sum_{m=1}^M V_m(x,y)$.  Hence, we can write}%Assuming that all distances between slices are the same, $\Delta_m = \Delta$ and $T = M \Delta$, we get
    \begin{equation}\label{eq: total potential approx}
    \begin{aligned}
     V_{total} \approx \sum_{m=1}^M \sum_{j=1}^N \sum_{k=1}^N V_m\left(j\frac{a}{N},k\frac{b}{N}\right) \frac{ab}{N^2}
     = \frac{ab}{N^2} \sum_{m=1}^M V_m, 
     % = \frac{ab T }{M N^2} \sum_{m=1}^M V_m
    \end{aligned}    
    \end{equation}
    where $a, b$ and $c$ are the height, width and thickness of the unit cell in {\AA}ngstrom. 
    }
    \ABnew{With the unit cell volume $\Omega = abc$ in cubic {\AA}ngstrom, the MIP is
    \begin{equation}\label{eq: average potetial}
        V_0 = \frac{1}{\Omega} V_{\text{total}} = \frac{1}{N^2c} \sum_{m=1}^M V_m.
    \end{equation}
    }
     % where $abc$ is the unit cell volume in cubic Angstrom}
    % The sum $\sum_{m=1}^M V_m$ approximates the total potential, namely
    % \begin{equation*}
    % \begin{aligned}
    % V_{total} & = \int_z \int_y \int_x V(x,y,z)\mrm{d}_x\mrm{d}_y\mrm{d}_z 
    % %= \sum_{m=1}^M \int_y \int_x V_m(x,y) \mrm{d}_x\mrm{d}_y \\
    % \end{aligned}
    % \end{equation*}
    % by sum of discrete grid with pixel size parameters. \OM{CONFUSED BY THIS PART. WHY DO WE GET MULTIPLIER IN FROM OF $V_{total}$?} \ABnew{Did I address your concern?} From the property of Riemannian sum we have
    % \begin{equation*}
    % \begin{aligned}
    % \sum_{m=1}^M \sum_{j=1}^N \sum_{k=1}^N V\left(j\frac{a}{N},k\frac{b}{N}, \Delta_m\right) \frac{abT}{N^2}\approx V_{total},
    % \end{aligned}    
    % \end{equation*}
    % here the $\Delta_m$ is the Fresnel propagation distance between each slice, $abc$ are the cubic volume of unit cell in Angstrom, and thickness of the specimen \ABnew{$T = cR$} is given by the unit cell's thickness scaled by a repeated factor $R$. \ABnew{Hence, the summation can be written as $V_{total}\frac{N^2}{abT} = V_0 \frac{N^2}{R},$ where $V_0 = \frac{ V_{total}}{abc}$} 
    
    %\OM{OM: here one should get $1/N^2$ since we approximate the integral with sums.}
    %\ABnew{AB: I need to update, since the discrete for realspace given by pixel size $dim (angstrom)/pixel$, we automatically get mean inner potential}
    \OM{By combining equations~\eqref{eq:det_phase} and \eqref{eq: average potetial}, we obtain 
    \[
    \log \det\left(\mbf{\hat S}\right)
        = -i \pi \lambda   Q T + i \sigma {N^2c}V_{0} + 2 \pi i n 
    \]
    and rearranging yields the estimate in equation~\eqref{eq:det_mip}.
    }

\OM{
\subsection{Mean Inner Potential and Unwrapping.}
An alternative way to compute the \ac{MIP} is by directly using the slices $\mbf O_m$. From equation~\eqref{eq:potential}, we observe that the phases of $\mbf O_m$ are $\sigma V_m(x,y)$ wrapped periodically due to the complex exponent. Thus, $\sigma V_m(x,y)$ can be recovered from $\mbf O_m$ by phase unwrapping\cite{herraez2002fast} up to an additive constant $c_m \in \mathbb R$,
\[
\text{\textbf{Unwrapp}}(\mbf O_m) = \sigma V_m(x,y) + c_m.
\]
In crystallographic applications, the constant can be further estimated by assuming that the most prevalent part of the observed object is vacuum, which does not cause scattering. Hence, the minimal value of the potential is almost zero, 
% $\textrm{Mode}(V_m) = 0$ 
leading to
\[
c_m = \min_{x,y} \text{\textbf{Unwrapp}}(\mbf O_m)_{x,y}.
\]
and
\[
V_m(x,y) = \frac{1}{\sigma} \left( \text{\textbf{Unwrapp}}(\mbf O_m) - c_m \right).
\]
We can then use the above values to compute the total inner potential via equation~\eqref{eq: total potential approx}.
} 

\subsection{Transmission Matrix Reconstruction.}
In this section, we describe the approach used to estimate the transmission matrix from experimental 4D-STEM data of a MoS$_2$ crystal. Let $\mbf{I} \in \R^{S_y \times S_x \times N \times N}$ denote the collection of $N \times N$ diffraction patterns obtained by illuminating the crystal with focused electron beams for a total $S = S_y \times S_x$ scanned specimen positions. Flattening the dataset as $\mbf{\hat I} \in \R^{N^2 \times S}$ and combining the probes as $\mbf{P} \in \C^{N^2 \times S}$ allows us to rewrite the reconstruction as a matrix estimation problem
\begin{equation}
\begin{aligned}
&\underset{\mbf{O}_m , m\in[M]}{\text{minimize}}
& & \norm{\sqrt{\mbf{\hat I}} - \card{\mathbf{F}_{2D}\mathbf{A}\mathbf{P} } }^2_F \\
& \text{subject to}
& & \mathbf{A} = \prod_{m= 1}^{M}\mathbf{G}_{m}\mbf{O}_m,
\label{Eq:Opt_Prob}
\end{aligned} 
\end{equation}
where the constraint is precisely equation~\eqref{eq: transmission matrix}. To solve equation~\eqref{Eq:Opt_Prob}, we employ the sparse matrix decomposition algorithm \cite{bangun2022inverse} which alternates between minimizing the objective for arbitrary $\mbf A$ and projecting it to satisfy equation~\eqref{eq: transmission matrix}. \ABnew{The reconstruction is performed with $3$ slices with a total thickness of approximately $4.6722$~nm. Since we have a total of $128 \times 128$ scan pixels with a distance between adjacent scan positions of $0.026$~nm/pixel, the scan area results to $3.328 \times 3.328$~nm$^2$. These parameters are used to calculate the \ac{MIP} of MoS$_2$ form experimental data given in Table~\ref{tab:mip} as well as the projected potential in Fig.~\ref{fig:recon_mip}d.}

% There are many approaches to solve the optimization problem \eqref{Eq:Opt_Prob}. The diagonal structure of matrix $\mbf{O}_m$ for each slice $m$ allows us to implement for instance sparse matrix decomposition algorithm, as discussed in \cite{bangun2022inverse}.
%{Some text moved up}
%Additionally, since phase factor is periodic up to $2\pi n$ for integer $n$ the solution for estimating mean inner potential should be scaled with modulo $2\pi$. 
%}
%\A{The estimation of {the \ac{MIP}} from {the} determinant for simulated and experimental data is discussed in Section \ref{sec:numerics}.}

\section{Data availability} \label{sec:datasets}
The data used to produce the results presented in this paper are available at Zenodo: \href{https://zenodo.org/records/14646184?token=eyJhbGciOiJIUzUxMiJ9.eyJpZCI6ImViZDlmM2FjLTUyNjItNDYzMi05YzYzLTRhNDQzNTI4YzZjZiIsImRhdGEiOnt9LCJyYW5kb20iOiJhZTFhZjdlOWYxMzNjOWZiMzkzMWI0MDU2MzA5YzhjYSJ9.Px6wYPGRKvFcNYnzfA_KBkG0SUspPeBE38Obfr--7IrsgxCjyugcqA5J0PrtPg_egIMR-wCMYo70Jl_BO4somQ}{https://zenodo.org/records/14646184}. 
In the following, we give the parameters of both simulated and experimental datasets.

\subsection{Simulated Datasets.}
Three crystalline materials are used:
%We describe the parameters used to generate simulated datasets for three crystalline materials: 
gallium arsenide (GaAs), a binary compound semiconductor, strontium titanate (SrTiO$_3$), a ternary oxide, and gold (Au), a single element crystal. 

\begin{table*}[ht!]
  \caption{Simulation parameters used to generate scattering and transmission matrix with both the Bloch wave and the multislice method, respectively. Structure data for GaAs \cite{wyckoff1963interscience}, SrTiO$_3$ \cite{mitchell2000crystal} and Au \cite{wyckoff1963interscience} was used. In the multislice data, each slice comprises the projected potential of all atoms within a single atomic layer. The Fresnel distance is the distance which the projected potential of a single slice is required to be propagated and corresponds to the lattice spacing along \hkl[001].}
\label{tab:params_multislice}
  \small
  \centering
  \begin{tabular}{llll}
    \toprule
     \bfseries Parameters & \bfseries \bfseries GaAs (F$\overline{\mathbf{4}}$3m)& \bfseries SrTiO$_3$ (Pm$\overline{\mathbf{3}}$m) & \bfseries Au (Fm$\overline{\mathbf{3}}$m)\\
     \midrule
      Unit cell dimensions (a,b,c) (\AA) & $(5.6533, 5.6533, 5.6533)$ & $(3.905, 3.905, 3.905)$ & $(4.08, 4.08, 4.08)$ \\
      {Unit cell angles ($\alpha,\beta,\gamma$) (°)} & \BM{$(90, 90, 90)$} & \BM{$(90, 90, 90)$} & $(90, 90, 90)$\\
      {Pixel size }(\AA)& $(0.1739,0.1739) $& $(0.1201,0.1201)$& $(0.1255,0.1255)$\\
      Fresnel distance (\AA) & $(1.413325, 1.413325)$ & $(1.9525, 1.9525)$ & $(2.04, 2.04)$\\
      e\textsuperscript{-} acceleration voltage (keV) & 300 & 300 & 300\\
    \bottomrule
  \end{tabular}
\end{table*} 

These materials differ in terms of total number and type (atomic number Z) of elements, and cubic unit cell dimension. Representations of the unit cells %of each material in projection along the \hkl[001] crystallographic direction and in perspective view 
are shown in Fig.~\ref{fig:materials}.
The parameters used to generate the simulated datasets are summarized in Table~\ref{tab:params_multislice}, outlining crucial characteristics such as unit cell dimensions, pixel size, and Fresnel distance for each material. These parameters serve as input for the simulation of the electrostatic potentials and the generation of the structure matrices. %, both of which are fundamental for understanding the scattering behavior of these materials when subjected to electron beams. 
%\ABnew{Done}
%\OM{I would remove in the previous sentence the part about starting from", both of which ..." We said it many times already, there is no point in repeating it again.} 

Our implementation of the multislice and the Bloch wave method, as well as the parameters used to generate the transmission and the scattering matrices are based on the MATLAB code by Durham et. al., 2022 \cite{durham2022accurate}.
\begin{table}[hbt!]
  \caption{Experimental 4D-STEM data acquisition parameters for MoS$_2$ %\BMnew{shall we move this table in the appendix or SI?} 
  %\ABnew{But we dont have any information related to the experimental data?}
  }
\label{tab:params_stem}
  \small
  \centering
  \begin{tabular*}{\linewidth}{ll}
    \toprule
     \bfseries Parameters & 
     \bfseries \bfseries Data  \\
      \midrule
      %Microscope & Hitachi HF5000\\
      %Detector  & Merlin4EM Medipix3\\
      Acceleration voltage (keV) & $200$\\
      Convergence semi-angle (mrad) & $32$\\
      Scan dimension (pixel) & $128\times128$\\
      Diffraction pattern dimension (pixel) & $256\times256$\\
      Scan rotation & $176^\circ$\\
      Real space pixel size (\AA) & $0.26$\\
    \bottomrule
  \end{tabular*}
\end{table}  

\subsection{Experimental Dataset.}
For the experimental process, 2H-MoS$_2$ sheets were exfoliated from a bulk crystal using a poly-dimethylsiloxane elastomer film placed on a glass slide and subsequently transferred onto a holey silicon nitride membrane for \ac{TEM} analysis. 
The experimental data was collected using a probe-corrected Hitachi HF5000 field emission microscope in scanning \ac{TEM} mode (STEM), operating at an acceleration voltage of $200$ keV with a beam current of approximately $7.4$~pA. The intensities of the diffraction patterns were recorded with a Medipix3 Merlin4EM camera, with a resolution of $256 \times 256$ pixels. The distance between adjacent scan points was set to $26$ pm in both the horizontal (x) and vertical (y) directions for the acquisition of an array of $128 \times 128$ diffraction patterns. Each diffraction pattern was acquired with $6$-bit dynamic range within $0.5$~ms.

\section*{Code availability} 
The source code to reproduce the result in this paper is available at \url{https://github.com/bangunarya/eigenstructure_bloch_ms} and \url{https://github.com/bangunarya/inverse_multislice}. 

\thispagestyle{empty}
 
\bibliography{references}
 
\section*{Acknowledgements}% (not compulsory)}

%Acknowledgements should be brief, and should not include thanks to anonymous referees and editors, or effusive comments. Grant or contribution numbers may be acknowledged.
\BMnew{We would like to thank Prof. K.-M. Caspary for making possible the 4D-STEM experiment at ER-C-1, Forschungszentrum Jülich.}

\section*{Author contributions statement}

\BMnew{A.B. conceptualized the study, wrote the original draft of the paper, implemented the Python code for generating and eigenstructure analysis of transmission and scattering matrix, as well as for simulating diffraction data. O.M. \ABnew{developed theoretical bound of differences between transmission and scattering matrix, as well as contributed to the mean inner potential (MIP) formulation for both logarithmic and unwrap version.} B.M. conducted the 4D-STEM experiment and contributed to the code review, data analysis and interpretation. All authors discussed the results, wrote and reviewed the manuscript.}

\cleardoublepage
\appendix
\section{Supplementary Information}
%This section is presented as supplement materials for the main article.
We describe the original multislice formulation, representation of the Fourier transform as a matrix, as well as additional numerical experiments.

\subsection{Multislice Method.}  
We present the classical formulation of the multislice method and briefly present the derivation of an alternative formulation based on matrices. The complete derivation is given in Bangun et. al., 2022 \cite{bangun2022inverse}.

Suppose we have a matrix that represents the $m-$th slice of the electrostatic potential of the specimen for a thin slice, similar to equation~\eqref{eq:potential}.
$$
\exp\left(i\sigma \int_{z_m}^{z_{m +1}} V(x,y,z) dz\right)
$$
In fact, for each combination axis $x,y$ we collect the potential and construct it into matrix $\mbf{X}_m \in \C^{N \times N}$ for a total slice number of $m \in [M]$. The interaction between the first slice and the electron wave for focused beam illumination at a specific scanning point $s$ is given by
$$
\mbf{E}_1 = \mbf{X}_1 \circ \mbf{P}_s
\in \C^{N \times N}$$
The exit wave is used as the source for the next slice by propagation using the Fresnel propagator $\mcl{V}$,
$$
\mbf{E}_m = \mcl{V}\left(\mbf{E}_{m-1} \right) \circ \mbf{X}_{m} \in \C^{N \times N}
\quad \text{for} \quad m \in \{2,3,\dots,M \}.
$$
%\OM{The detail?}
The Fresnel propagator is given in Fourier space for a specific distance between slices. % \OM{Why suppose?} 
We define the Fourier space of the Fresnel propagator matrix $\mbf{H}_m \in \C^{N \times N}$. %where i
Its elements are given in equation~\eqref{eq:fresnel_element}. Hence, the Fresnel propagation can be written as
$$
\mcl{V}\left(\mbf{E}_{m}\right) := \mcl{F}^{-1}\left(\mbf{H}_m \circ \mcl{F}\left(\mbf{E}_m\right) \right).
$$
As a result, the intensity recorded at a specific scanning point $s$ in the microscope's detector plane is given as
$$\mbf{I}_s  = \card{\mcl{F}\left( \mbf{E}_M\right)}^2  \in \R^{N \times N},$$
where the absolute value is applied to each element of the matrix.

It can be seen that, a separation of the entanglement between the electron wave and the object is difficult. Hence, an exact representation of the transmission function of the object, independent of the propagated exit wave starting from the probe illumination, is not available. As discussed in Bangun et. al., 2022 \cite{bangun2022inverse}, the reformulation by leveraging the property of element-wise or Hadamard matrix product, namely
\begin{equation}
\begin{aligned}
\vct{\mbf A \circ \mbf B} &= \diag{\vct{\mbf A}} \vct{\mbf B}\\
&= \diag{\vct{\mbf B}}  \vct{\mbf A},
\end{aligned}
\label{eq:haddamard}
\end{equation}
where the function $\text{vec}:\C^{N\times N} \mapsto \C^{N^2}$ is used to convert a matrix into a vector and $\text{diag}:\C^{N^2} \mapsto \C^{N^2 \times N^2}$ is the function to construct a diagonal matrix from a vector. The vectorized intensity at scanning point $s$ can be written as
$$
\mbf{i}_s = \card{\mbf{F}_{2D} \left( \prod_{m=1}^M \mbf{G}_m \mbf{O_m}  \right) \mbf{p}_s}^2 \in \R^{N^2} \,\text{for}\, s \in [S]
$$
Using this formulation, we can disentangle the illuminating electron wave and the object in order to get a pure transmission function 
$$
\mbf A = \prod_{m=1}^M \mbf{G}_m \mbf{O_m},
$$
where the formulation of Fresnel matrix $\mbf{G}_m$ is derived in equation~\eqref{eq:fresnel_matrix}.

\begin{figure}[htb!]
    \centering
    \includegraphics[width=\columnwidth]{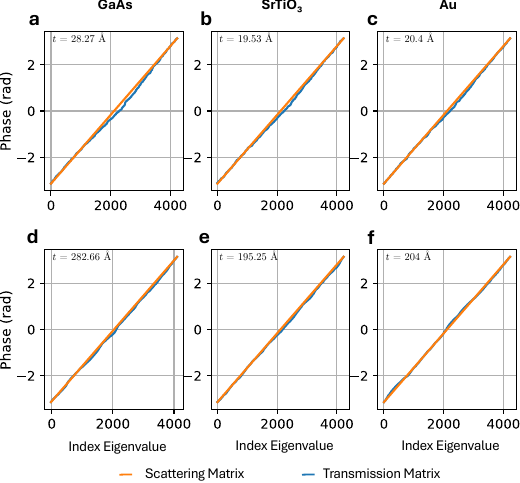}
    \caption{Phase distribution of eigenvalues obtained from transmission matrix and scattering matrix for two different thicknesses $T$. \textbf{a}, GaAs of $T=28.27$\,\AA{} (SD~$=0.098$~rad). \textbf{d}, GaAs of $T=282.66$\,\AA{} (SD~$=0.043$~rad). \textbf{b}, SrTiO$_3$ of $T=19.525$\,\AA{} (SD~$=0.068$~rad). \textbf{e}, SrTiO$_3$ of $T=195.25$\,\AA{} ({SD}~$=0.0384$). \textbf{c}, Au of $T=20.4$\,\AA{} (SD~$=0.06$~rad). \textbf{f}, Au of $T=204$\,\AA{} (SD~$=0.05$~rad).}
    \label{fig:eig_dist_appendix}
\end{figure}

\begin{figure*}[htb!]
    \centering
    \includegraphics[width=\textwidth]{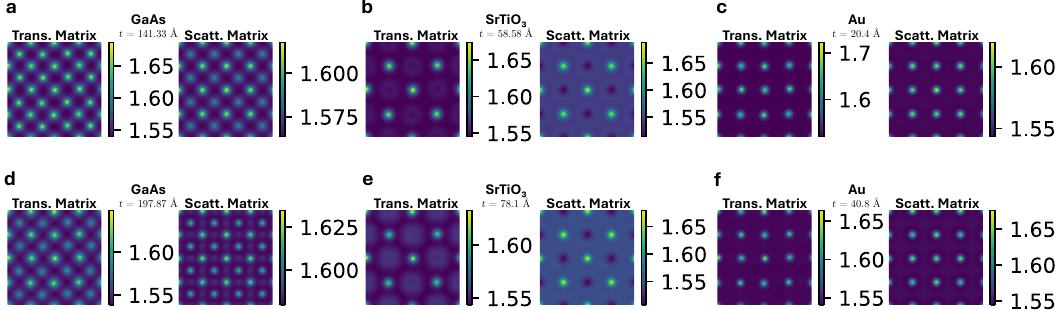}
    \caption{Projected potential of $2\times2$ unit cells in \hkl[001] zone axis using the transmission matrix (1\textsuperscript{st}, 3\textsuperscript{rd} and 5\textsuperscript{th} column) and the scattering matrix (2\textsuperscript{nd}, 4\textsuperscript{th} and last column) for different thicknesses $t$. \textbf{a}, GaAs with thickness $141.33$\,\AA{} (25 unit cells) and, \textbf{d}, $197.87$\,\AA{} (35 unit cells). \textbf{b}, SrTiO$_3$ with thickness $58.58$\,\AA{} (15 unit cells) and, \textbf{e}, $78.1$\,\AA{} (20 unit cells). \textbf{c}, Au with thickness $20.4$\,\AA{} (5 unit cells) and, \textbf{f}, $40.8$\,\AA{} (10 unit cells).% Left figure is calculated with the transmission matrix, and the right figure with the scattering matrix.
    }
\label{fig:proj_potential_appendix}
\end{figure*}

\begin{figure}[htb!]
    \centering
    \includegraphics[width=\columnwidth]{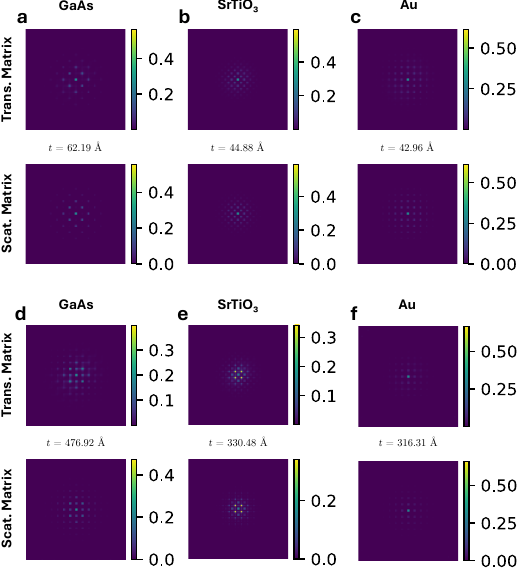}
    \caption{Diffraction patterns along the \hkl[001] zone axis calculated using the transmission matrix (1\textsuperscript{st} and 3\textsuperscript{rd} row) and the scattering matrix (2\textsuperscript{nd} and last row) for different thicknesses $T$. \textbf{a}, GaAs with  $T = 62.1863$\,\AA\, and, \textbf{d},  $T = 476.9173$\,\AA. \textbf{b}, SrTiO$_3$ with simulated thickness $T = 44.88$\,\AA \,and, \textbf{e}, $T = 330.48$\,\AA. \textbf{c}, Au with simulated thickness $T = 42.955$\,\AA \,and, \textbf{f},  $T = 316.305$\,\AA.}
\label{fig:bragg_images_appendix}
\end{figure}  

\subsection{Matrix Representation of Fourier Transforms.}
We present a matrix representation of Fourier transforms. %For t
The one-dimensional Fourier transform can be written as a matrix-vector product.
Here, the term one-dimensional Fourier matrix represents the discrete implementation of the Fourier basis, i.e., when we sample the Fourier basis and store it as a matrix $\mbf{F}_{1D} \in \C^{N \times N}$. Hence, the one-dimensional Fourier transform of vector $\mbf x \in \C^{N \times N}$ can be written as
$$
\mbf{y} = \mbf{F}_{1D} \mbf{x},
$$
where the one-dimensional Fourier transform is given by \cite[eq.\,5.44]{jain1989fundamentals},

$$
\mathbf{F}_{1D} = \frac{1}{\sqrt{N}}\begin{pmatrix}
  e^{\frac{-i2\pi f_1x_1}{N}} \quad e^{\frac{-i2\pi f_1x_2}{N}}&\dots& e^{\frac{-i2\pi f_1x_N}{N}} \\
  e^{\frac{-i2\pi f_2x_1}{N}} \quad e^{\frac{-i2\pi f_2x_2}{N}}&\dots& e^{\frac{-i2\pi f_2x_N}{N}} \\
  \vdots&  \dots&\vdots\\
  e^{\frac{-i2\pi f_Nx_1}{N}} \quad e^{\frac{-i2\pi f_Nx_2}{N}}&\dots& e^{\frac{-i2\pi f_Nx_N} {N}},\\
\end{pmatrix} \in \C^{N \times N}.
$$

Here, $x_j$ for $j \in [N]$ represents the sampling points at %on the 
evenly spaced real space coordinates, and $f_j$ for $j \in [N]$ is the sample in the Fourier space. Along the same line, the two-dimensional Fourier matrix can be constructed using the Kronecker product between two of the one-dimensional Fourier matrices $\mbf{F}_{2D} = \mbf{F}_{1D}\otimes \mbf{F}_{1D} \in \C^{N^2 \times N^2}$. The two-dimensional Fourier transform of matrix $\mbf{X} \in \C^{N \times N}$ can be written as  $
\mbf{F}_{2D} \text{vec}\left(\mbf X\right) \in \C^{N^2}
$.
\subsection{Additional Numerical Experiments.}
In order to highlight the consistency of results that are obtained using scattering and transmission matrix, when applied to data of specimens with highly different thickness, we present additional numerical experiments, that are, diffraction patterns shown in Fig.~\ref{fig:bragg_images_appendix}, phase of eigenvalues as in Fig.~\ref{fig:eig_dist_appendix}, as well as projected potential as in Fig.~\ref{fig:proj_potential_appendix}.

\end{document}